%% file: nussdpnu.tex
\documentstyle[12pt,epsf]{article}
\makeatletter
\def\@maketitle{%
  \newpage
  \null
 \vspace{-30mm}         
 \begin{flushright}
 \prepno\par
 \prepdate
\end{flushright}
  \vskip 2em%
  \begin{center}%
  \let \footnote \thanks
    {\LARGE \@title \par}%
    \vskip 1.5em%
    {\large
      \lineskip .5em%
      \begin{tabular}[t]{c}%
        \@author
      \end{tabular}\par}%
  \end{center}%
  \par
  \vskip 1.5em}
\makeatother
\begin{document}
\setcounter{section}{-1}
\renewcommand{\theequation}{\thesection.\arabic{equation}}
\newcommand{\my}{\setcounter{equation}{0}}
\def\prepno{DPNU-98-07 \\ hep-th/9802037}%
\def\prepdate{January 1998}

\include{absdpnu}

\tableofcontents
\newpage
\input{dpnu970}

\input{dpnu971}

\input{dpnu972}
\input{dpnu973}

\input{dpnu974}

\input{dpnu975}

\input{dpnu976}

\input{dpnu977}

\input{dpnuref}

\end{document}

%% file: absdpnu.tex
\title{
 Zero-Mode Problem on the
 Light Front
 \thanks
 {Lectures given at 10 th Annual Summer School and Symposium on
 Nuclear Physics (NUSS 97), ``QCD, Lightcone Physics
 and Hadron Phenomenology'', Seoul, Korea,
 June 23-28, 1997.}
\author{ KOICHI \hspace{0.22cm} YAMAWAKI
\thanks{
Email address: yamawaki@eken.phys.nagoya-u.ac.jp}
\\
Department of Physics\\
 Nagoya University \\
  Nagoya,   464-01 \\
  Japan }
  }
\maketitle
\begin{abstract}
A series of lectures are given to discuss the zero-mode problem 
on the light-front (LF) quantization  with
special emphasis on the peculiar realization of the
trivial vacuum, the spontaneous symmetry breaking (SSB)
and the Lorentz invariance.
We first identify the zero-mode problem on the LF.
We then discuss {\it Discrete Light-Cone Quantization (DLCQ)} which was first 
introduced by Maskawa and Yamawaki (MY) to solve the zero-mode problem
and later advocated by Pauli and Brodsky in a different  context.
Following MY, we present 
canonical formalism of DLCQ and  
the {\it zero-mode constraint}
 through which the zero mode can actually be solved away 
in terms of other modes, thus {\it establishing the trivial vacuum}.
Due to this trivial vacuum, 
existence of the massless Nambu-Goldstone (NG) boson  
coupled to the current is guaranteed 
by  the {\it non-conserved charge} 
such that $Q\vert 0\rangle=0$ and $\dot{Q}\ne 0$
but {\it not} by the 
{\it NG theorem}
which in the equal-time quantization 
ensures existence of the massless NG boson 
coupled to the current with the charge  
$Q\vert 0\rangle \ne 0$ and $\dot{Q} =0$. 
The SSB (NG phase) in DLCQ can be 
realized on the trivial vacuum only when an 
explicit symmetry-breaking mass of the NG boson $m_{\pi}$  
is introduced 
so that the NG-boson zero mode integrated over the LF 
exhibits {\it singular behavior} $ \sim 1/m_{\pi}^2$ 
in such a way that $\dot{Q}\ne 0$  
in the symmetric limit $m_{\pi}\rightarrow 0$. 
We also demonstrate this realization more explicitly in a concrete model, 
the linear sigma model,
where the role of zero-mode constraint is clarified. 
We further point out, in disagreement with Wilson et al., that 
for {\it SSB in the continuum LF theory}, 
the {\it trivial vacuum collapses} 
due to the special nature of the zero mode which is no longer 
the problem of a single mode $P^+\equiv 0$ but of 
the  accumulating point $P^+\rightarrow 0$, in sharp contrast to DLCQ. 
Finally, we  discuss 
the no-go theorem
 of Nakanishi and Yamawaki, which 
 forbids exact LF restriction 
 of the field theory that satisfies the Wightman axioms.
The well-defined LF theory exists only 
at the sacrifice of the Lorentz
invariance. Thus DLCQ as well as any other regularization
on the exact LF has 
{\it no  Lorentz-invariant limit 
as the theory itself}, although 
we can argue, based on an {\it explicit solution} of the
dynamics (i.e., perturbation),
that the {\it Lorentz-invariant limit} can be realized
on the {\it c-number quantity} 
like S matrix which has {\it no reference to the fixed LF}.
\end{abstract}


%% file: dpnu970.tex
\section{Introduction}

Much attention has recently been paid to the light-front (LF) 
quantization\cite{Dira} as a promising approach to solve the
non-perturbative dynamics \cite{Wils,Yama,review}. 
The most important aspect of the LF quantization
is that the physical LF vacuum is simple, or even trivial \cite{LKS}.
However, such a trivial vacuum, which is vital to the whole LF approach, 
can be realized only if we can 
remove the so-called zero mode out of the physical Fock space 
(``zero mode problem''\cite{MY}).
  
Actually, the Discrete Light-Cone Quantization (DLCQ)\footnote{
~The name ``light-cone quantization'' is actually confusing, since it 
is not on the
light cone but on the light front which agrees with the former 
only in $(1+1)$ dimensions. However,
here we simply follow the conventional naming of the majority of 
the literature.}  was first introduced 
by Maskawa and Yamawaki (MY)\cite{MY}  in 1976 \footnote{
~The DLCQ was also considered by Casher\cite{casher}
independently in a somewhat different context.  
}
to resolve the zero mode
problem and was advocated by Pauli and
Brodsky\cite{PB} in 1985 in a different context.
The zero mode in DLCQ is clearly isolated from other modes and hence
can be treated in a well-defined manner without ambiguity, 
in sharp contrast to
the continuum theory where the zero mode is the accumulating point and hard to
be controlled in isolation \cite{NY,TY}.  
In DLCQ, MY\cite{MY} in fact discovered a
constraint equation for the zero mode (``{\it zero-mode constraint}'')
 through which 
the zero mode becomes dependent on other
modes and then they observed that the zero mode can be removed from 
the physical
Fock space by solving the zero-mode constraint, thus {\it establishing the 
trivial LF vacuum in DLCQ}. 
 
Such a trivial vacuum, on the other hand, might confront the usual picture of
complicated non-perturbative vacuum structure 
in the equal-time quantization corresponding to   
confinement, spontaneous 
symmetry breaking (SSB), etc.. Since the vacuum is proved trivial in 
DLCQ \cite{MY},
the only possibility to realize such phenomena would be through
the complicated structure of the operator  
and only such an operator would be the zero mode.
In fact the zero-mode constraint implies that 
the zero mode carries essential information on the complicated dynamics.
One might thus expect that explicit solution of the zero-mode constraint
in DLCQ would give rise to the SSB 
while preserving the trivial LF vacuum.
Actually, several authors have argued in (1+1)
dimensional models that the solution to the zero-mode constraint
might induce spontaneous
breaking of the discrete symmetries \cite{HKSW,Rob,BPV}. However,
the most outstanding feature of the SSB is
the existence of the Nambu-Goldstone (NG) boson for the continuous symmetry
breaking in (3+1) dimensions.
 
In recent works done in collaboration with Yoonbai Kim
and Sho Tsujimaru \cite{KTY,TY}, we in fact found,
within the canonical formalism of DLCQ\cite{MY}, how the SSB
manifests itself on LF in (3+1) dimensions through existence of the NG 
boson while keeping the vacuum trivial. \footnote{
~Actually, the language of ``SSB'' is somewhat confusion
on the LF as we will see later and we will use the word
``Nambu-Goldstone (NG) phase'' instead.
}

In the above works\cite{KTY,TY} we encountered a striking feature of the
zero mode of the NG boson: {\it Naive use of the zero-mode constraint
does not lead to the NG phase at all},
with the NG boson being totally decoupled ((false) ``no-go theorem''). 
It implies that the above approach\cite{HKSW,Rob,BPV}, as it stands,
fails to realize the NG phase.
Indeed it is inevitable to introduce an infrared regularization through
explicit-breaking mass of the NG boson $m_\pi$. 
The NG phase can only be realized via peculiar
behavior of the zero mode of the NG-boson field:
{\it The NG-boson zero mode, when integrated over the LF, 
must have a singular behavior $\sim 1/m_{\pi}^2$} in the symmetric limit
$m_\pi^2 \rightarrow 0$.
This was demonstrated both in
a general framework of the LSZ reduction formula  
and in a concrete field theoretical model, 
the linear sigma model.
The NG phase is in fact realized in such a way that the 
{\it vacuum is trivial}, $Q\vert0\rangle =0$,
while {\it the LF charge $Q$ is not conserved}, $\dot{Q} \ne 0$, 
even in the symmetric limit 
$m_\pi^2 \rightarrow 0$.

Note that this is not an artifact of DLCQ 
but is a general feature of NG
phase realized on the trivial vacuum of LF: 
As far as the vacuum is trivial and
hence symmetric, the only possibility to realize the non-symmetric
physical world is that the Hamiltonian is {\it not symmetric at quantum level}
(if both vacuum and Hamiltonian were symmetric, then the physical world would 
also be symmetric). It might look like the usual explicit symmetry 
breaking but is essentially different in the sense 
that we do have massless NG boson coupled to the current.
The existence of massless NG boson
coupled to the current is in fact
ensured by the charge $Q\vert0\rangle =0$ and $\dot{Q} \ne 0$,
which is quite opposite to the NG theorem in the equal-time quantization with the
charge $Q\vert0|\rangle\ne 0$ and $\dot{Q}=0$. 
The above singular behavior of the NG-boson
zero mode is thus a manifestation of the same 
SSB phenomenon as that dictated by the NG theorem of the
equal-time quantization, though 
in a way quite different than the NG theorem.
 
It is rather remarkable that in the continuum theory the
 SSB cannot be formulated consistently  with the trivial vacuum \cite{TY} 
in sharp contrast to DLCQ. 
It was demonstrated\cite{TY}
with careful treatment of the boundary condition (B.C.)
that as far as the sign function is
used for the standard canonical LF commutator, the {\it LF charge
does not annihilate the vacuum in the continuum LF theory}
in disagreement with Wilson et al.\cite{Wils}. 
This in fact reflects the problem of the zero mode in the 
continuum theory: The problem is {\it not a single mode} $P^+\equiv 0$ 
with just measure zero
but the {\it accumulating point} $P^+\rightarrow 0$ which
{\it cannot be removed} by simply dropping the exact zero mode $P^+\equiv
0$ \cite{NY}. 

The zero-mode problem in this sense is more serious in the context of Lorentz invariance.
There in fact exists a {\it no-go theorem}
 found by Nakanishi and Yamawaki\cite{NY}
which {\it forbids the exact LF restriction}
 of the field theory that satisfies the 
Wightman axioms, due to the zero mode as the accumulating point: 
In particular, {\it even a free theory does not exist} on the exact
LF. This implies that in order to make the theory well-defined on the
exact LF, we are forced to give up some of the Wightman axioms, most naturally
the Lorentz invariance. Indeed, the DLCQ and the ''$\nu$-theory''\cite{NY}, 
defined on the exact
LF, are  such theories: The  theory itself  (operator,
Hilbert space, etc) explicitly violates
the Lorentz invariance and never recovers it even in the continuum 
limit of $L \rightarrow \infty$ ($\nu \rightarrow 0$ limit in the
$\nu$-theory). 
It is actually at the sacrifice of the Lorentz invariance 
that the trivial vacuum is realized in DLCQ and $\nu$-theory. However, {\it 
even though the 
theory itself does not recover 
the Lorentz invariance} in these theories,  
we still can argue \cite{NY,TY}, at least in the perturbation 
theory, that 
{\it the Lorentz invariance is realized on the
c-number quantity like the S matrix which has no reference to the fixed LF},
while keeping the vacuum polarization absent. Such a program is based on
the explicit solution of the dynamics, like perturbation. Actually, recovering 
the Lorentz invariance is highly dynamical issue in the LF quantization, the situation being 
somewhat analogous to the lattice gauge theories. The non-perturbative way to
recover the Lorentz invariance remains a big
challenge for any LF theory.
 
In this lecture we shall fully discuss the above result 
and the zero-mode problem
in the general context. The lecture is largely based on the above recent 
works\cite{KTY,TY} and also 
on the very old ones\cite{MY,NY} which are most relevant to the 
zero-mode problem.
 
In Section 1 we first explain a key feature of the LF quantization,
the trivial vacuum and physical Fock space. 
Then we discuss the subtlety of the zero mode 
which might jeopardize 
this nice property (zero-mode problem).
In Section 2 we give the canonical formalism of DLCQ through 
the Dirac's quantization for
the constrained system, with particular emphasis on the zero mode: Detailed derivation
of the LF commutator and the zero-mode constraint are given.
Discussions are also given to the subtlety 
of the B.C.'s of DLCQ 
in the continuum limit.   
In Section 3 we first discuss in general the peculiarity 
of the realization of the NG phase on the LF
by the charge $Q\vert0\rangle =0$ and $\dot{Q} \ne 0$. Then in DLCQ 
we show that if we were in the exact symmetric theory with the
NG-boson mass exactly zero from the beginning, 
the NG phase would never be realized on the LF: The NG boson simply decouples
even in the continuum limit of DLCQ ((false) ``no-go theorem''). 
We then show that such
an inconsistency in DLCQ is resolved by introducing an explicit-symmetry-breaking
mass of the NG boson $m_{\pi}$ in such a way that the NG-boson zero mode integrated over LF 
has a singular behavior $\sim 1/m_{\pi}^2$ in the symmetric limit
$m_\pi^2 \rightarrow 0$, in accordance with the above general statement 
$Q\vert0\rangle =0$ and $\dot{Q} \ne 0$. 
In Section 4 we demonstrate such a peculiar realization in a concrete model,
the linear sigma model, by solving the zero-mode constraint perturbatively. 
Section 5 is devoted to discussions on the zero-mode problem in the continuum 
theory: The LF vacuum cannot be trivial in the NG phase 
due to the peculiarity of the zero mode
in the continuum LF theory.
In Section 6 we discuss the problem of Lorentz invariance in view of the
no-go theorem which forbids the well-defined LF restriction due to
the peculiarity of the zero mode. A way to recover the Lorentz invariance
is suggested. 
Summary and discussions are given in Section 7.


%% file: dpnu971.tex

\section{Trivial Vacuum and Physical Fock Space}
\my
The light-front (LF) quantization is based on the field commutator 
on the LF which is the hypersurface
 tangent to the light cone (LF coincides with the light cone only in
(1+1) dimensions but not in (3+1) dimensions).
It is also  called null-plane 
quantization, lightlike quantization, front form, and 
light-cone quantization (even though in (3+1) dimensions),
etc.. Note that the LF quantization 
is inherent
in 
the {\it Minkowski space}, since the LF, 
the quantization plane,
is no longer meaningful in the Euclidean
space.
 
Throughout this lecture 
we use a convention of the LF coordinate: 
$x^{\mu}=(x^+,\vec{x})=(x^+, x^-, x^{\bot})$,
where 
\begin{eqnarray}
x^{\pm}= x_{\mp}\equiv\frac{1}{\sqrt 2}(x^0 \pm x^3),
\end{eqnarray}
with $x^+$ being the  ``time''  while $x^-$ the ``space'' coordinate
along the LF, and $
x^{\bot}=-x_{\bot} \equiv (x^1,x^2)$. 
The respective conjugate momenta $p_{\mu}= (p_+, p_-, p_{\bot})
=(p^-,\vec{p})$ are
\begin{eqnarray}
p_{\pm}=p^{\mp} \equiv\frac{1}{\sqrt 2}(p^0 \mp p^3),
\end{eqnarray}
with $p^-$ being the energy while $p^+$ the momentum, and
$p_{\bot}= -p^{\bot} \equiv -(p^1,p^2)$.
The momentum operator $P^+$ is a translation operator along
$x^-$ direction and is free from dynamics ({\it kinematical operator}),
while the energy
operator $P^-$, the Hamiltonian, is the evolution operator of time $x^+$,
which carries a whole information of the dynamics (dynamical operator).
Then $P^+$ is {\it conserved even in virtual processes}, while $P^-$ is
not.

\subsection{$P^+ >0$ and Trivial Vacuum/Physical Fock Space} 
Now, the most important feature of the LF quantization follows from
the spectral condition (see Fig.1): Both $P^+$ and $P^-$ as well as $P^0$ 
have positive-definite
spetra for the physical states\footnote{
~For the
moment we disregard the zero mode $P^+=0$ which is the central subject of
this paper and is to be fully discussed later.
}, namely, 
\begin{equation}
p^+,\quad p^- >0
\quad \Longleftrightarrow\quad
p^0 >0.
\label{sc}
\end{equation}
\begin{figure}[htbp]
\begin{center}
\leavevmode
\epsfxsize=5cm
\epsfbox{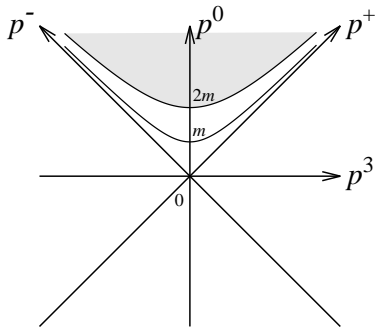}
\end{center}
\caption{The spectral condition (a single mass system).} 
\label{fig1:}
\end{figure}
 
This is contrasted to the equal-time case where 
only the energy operator possesses positive-definite spectrum, $p^0>0$, while
the momenta can take both positive and negative signs, 
$\mbox{\boldmath$p$}\equiv(p^1,p^2,p^3)>0 \quad {\rm and} <0$.
Actually, 
{\it all the characteristic features of LF quantization simply stem from the 
fact that  $P^+$ is a kinematical operator and at the same time
has positive-definite spectrum}.

First of all it implies that {\it the vacuum is trivial}, 
since we can always specify the eigenvalue of $P^+$ without
solving the dynamics and the vacuum $\vert 0\rangle$ 
can be defined kinematically as a state with the lowest eigenvalue
of $P^+$ without recourse to the $P^-$ spectrum; $P^+\vert 0\rangle =0$.
In view of the spectral condition (\ref{sc}) or Fig.1, we may infer that 
the trivial vacuum is also the true vacuum, namely,
\begin{eqnarray}
P^+\vert 0\rangle =0 \quad
\Longrightarrow \quad
P^-\vert 0\rangle =0.
\label{truev}
\end{eqnarray}
Upon such a trivial vacuum we can construct the {\it physical Fock space} 
in terms of
the {\it interacting Heisenberg field} 
whose Fourier components with $p^+>0$ and
$p^+<0$ are interpreted as the creation and 
annihilation operators, respectively \cite{LKS}.
This is also viewed as {\it absence of vacuum polarization}\cite{WIMF}, 
since all physical states including virtual states must have positive $P^+$ 
and at the same time respect conservation of $P^+$ as the kinematical 
operator (``momentum conservation'') even for the virtual states ($P^+>0$) 
which are to be communicated
to the vacuum ($P^+=0$) as the initial and final states in the 
vacuum polarization diagrams. 

To be more concrete, let us consider the self-interacting scalar theory in which
the interacting Heisenberg field $\phi(x)$ is Fourier-expanded as
\begin{equation}
\phi(x) = \frac{1}{(2\pi)^3}\int_0^{\infty} \frac{dp^+}{2 p^+} \int_{-\infty}^{\infty} d p^{\bot}
\left( a(\vec{p},x^+) e^{-i\vec{p}\vec{x}}+ a^\dagger(\vec{p},x^+)
e^{i\vec{p}\vec{x}} \right),  
\label{FT}
\end{equation}
where
we defined
\begin{eqnarray}
a(\vec{p},x^+)/2 p^+ \equiv \tilde{\phi}(\vec{p},x^+), \quad
a^\dagger(\vec{p},x^+)/2 p^+ \equiv \tilde{\phi}(-\vec{p},x^+)\quad (p^+ >0)
\end{eqnarray}
and $\vec{p}\vec{x}=p^+ x^- -p^\bot x^\bot$, with  $\vec{p} \equiv 
(p^+, -p^{\bot})$ 
being just momenta having nothing to
do with the dynamics. 

Now, the Heisenberg operators $a(x^+)$ and $a^\dagger(x^+)$, 
whose $x^+$-dependence is {\it not solved},  can be interpreted as the
creation and annihilation operators, respectively, as follows \cite{LKS}.
Let us assume  
translation invariance, namely existence of the operator $P^+$ such that 
\begin{equation}
[\phi(x),P^+]=i\partial_-\phi(x),
\label{trans}
\end{equation}
then it follows from (\ref{FT}):
\begin{equation}
[a(\vec{p},x^+), P^+] =  p^+ a(\vec{p},x^+), \quad
[a^\dagger(\vec{p},x^+),P^+] =- p^+ a^\dagger(\vec{p},x^+).
\label{cr-ann}
\end{equation}
Thus $a(x^+)$ and $a^\dagger(x^+)$ as they stand 
are lowering and raising operators of $P^+$ 
eigenvalue, respectively and the {\it spectral condition} $P^+ >0$
dictates existence of the {\it lowest eigenvalue 
 state} $\vert 0\rangle$ ($P^+\vert 0\rangle=0$),
the vacuum,  
 such that
\begin{equation}
a(\vec{p},x^+) \vert 0\rangle =0,
\label{trivac}
\end{equation}
which would coincide with the true vacuum from the spectral condition 
as in (\ref{truev}).
The physical Fock space may be constructed by operating the creation operators 
$a^\dagger$'s on the vacuum:
\begin{equation}
\vert \vec{p}_1,\vec{p}_2, \cdot\cdot\cdot, \vec{p}_n\rangle_{x^+}
= a^\dagger(\vec{p}_1,x^+) a^\dagger(\vec{p}_2,x^+)\cdot\cdot\cdot 
a^\dagger(\vec{p}_n,x^+) \vert 0 \rangle
\label{Fock}
\end{equation}
up to normalization.
On the other hand, the standard 
canonical commutator on LF reads (we shall discuss the derivation later):
\begin{eqnarray}
\left[ \phi(x), \phi(y)\right]_{x^+=y^+}= -\frac{i}{4}\epsilon (x^- -y^-)
\delta (x^\bot-y^\bot),
\label{CCR}
\end{eqnarray}
where $\epsilon (x^-)$ is the sign function:
\begin{eqnarray}
\epsilon (x^-) =\frac{i}{\pi} {\cal P}
\int_{-\infty}^{+\infty}\frac{dp^+}{p^+}
e^{-ip^+x^-}
&=&1 \quad (x^->0)\nonumber\\
&=&-1 \quad (x^- <0),
\label{sign}
\end{eqnarray}
with ${\cal P}$ being the principal value.
 From this and the inverse Fourier transform of (\ref{FT}),
\begin{eqnarray}
a(\vec{p},x^+) =
2 p^+
\int d\vec{x} e^{i\vec{p}\vec{x}}
\phi(x)
           =
           i
           \int d\vec{x} e^{i\vec{p}\vec{x}}
  \stackrel{\leftrightarrow}{\partial}_- \phi(x),
\label{invFT}
\end{eqnarray}
(similarly for $a^\dagger$),
we obtain 
the commutation relation of $a$ and $a^\dagger$ of the conventional
form:
\begin{eqnarray}
\left[a(\vec{p},x^+), a^\dagger(\vec{q},x^+)\right]
&=&
2 p^+ 2 q^+\int d \vec{x} d \vec{y} e^{i \vec{p}\vec{x}}
 \left[
\phi(x), \phi(y) \right]_{x^+=y^+}
e^{-i\vec{q}\vec{y}}
\nonumber \\
&=& (2\pi)^3 2 p^+ \delta (\vec{p}-\vec{q})
\label{hocr}.
\end{eqnarray}
Note again that the $x^+$-dependence of $a(x^+)$ and $a^\dagger(x^+)$ 
is {\it not} solved 
(they are Heisenberg fields) and whole procedure was done in a purely 
kinematical way without recourse to the dynamics (information of $P^-$). 

Here it is useful to compare the above feature of LF quantization 
with that of the equal-time quantization.
One might think the same type of arguments as those for 
the relation (\ref{cr-ann}) could be done for
the Fourier component with respect to 
momenta $\mbox{\boldmath $P$}\equiv (P^1,P^2,P^3)$ instead of $P^+$:
$
\phi(x) = \int \frac{d^3 \mbox{\boldmath$p$}}{(2\pi)^3 2 p^0}
\tilde{\phi}(\mbox{\boldmath $p$},x^0) 
e^{-i\mbox{\boldmath$p$}\mbox{\boldmath$x$}}.  
$
Alas, the spectral condition, $ 
\mbox{\boldmath $P$}\equiv (P^1,P^2,P^3)> 0$ and $<0$, 
which implies no lowest eigenvalue state for these operators
in sharp contrast to $P^+ >0$.
It is only $P^0$ that 
has positive-definite spectrum $P^0>0$ and  
a relation analogous to (\ref{cr-ann}) follows from 
\begin{equation}
[\phi(x),P^0]=i\partial_0\phi(x),
\end{equation}
instead of (\ref{trans}).
However, {\it specifying the definite eigenvalue} of $P^0$ for the Fourier modes
is {\it equivalent to solving the dynamics}
 at operator level, which is in general 
impossible except for the free theory which is completely solved. 
As everybody knows, in the free theory
 we in fact can construct the Fock space
based on the {\it explicit solution} of the free field equation of motion
(Klein-Gordon equation) for the Fourier modes
$\tilde{\phi}(\mbox{\boldmath $p$},x^0)$:
\begin{equation}
\ddot{\tilde{\phi}}(\mbox{\boldmath $p$},x^0) 
+ \omega^2 \tilde{\phi}(\mbox{\boldmath $p$},x^0) = 0 \qquad (\omega\equiv 
\sqrt{\mbox{\boldmath $p$}^2 +m^2}>0).
\end{equation}
This is nothing but the harmonic oscillator
and of course has two independent solutions:
\begin{eqnarray}
\tilde{\phi}(\mbox{\boldmath $p$},x^0) &=&\frac{1}{2\omega}
a(\mbox{\boldmath$p$})e^{-i\omega t} \quad (p^0=\omega),\nonumber\\
&=&\frac{1}{2\omega}
a^\dagger(-\mbox{\boldmath$p$}) e^{i\omega t} \quad (p^0=-\omega). 
\label{mode-free} 
\end{eqnarray}
These {\it solutions}  are in fact lowering and raising operators,
respectively, of the eigenvalue of $P^0$:
\begin{eqnarray}
[a(\mbox{\boldmath$p$}), P^0]
&=&i\dot{a}(\mbox{\boldmath$p$})
=-\omega a(\mbox{\boldmath$p$}), \nonumber\\
\left[a^\dagger(\mbox{\boldmath$p$}), P^0\right]
&=&i\dot{a}^\dagger(\mbox{\boldmath$p$})
=\omega a^\dagger(\mbox{\boldmath$p$}).
\end{eqnarray}
Then the spectral condition $P^0>0$ guarantees the existence of the
vacuum $\vert 0\rangle$ as the lowest eigenvalue state 
($P^0 \vert 0\rangle=0$) such that
$a(\mbox{\boldmath$p$}) \vert 0 \rangle=0$
{\it as the explicit solution of the dynamics}. 

\subsection{Advantages of the Light-Front Quantization}

Having discussed that the physical Fock space can be constructed 
upon the trivial vacuum even for the interacting Heisenberg field,
we may next discuss possible advantages of LF quantization over
the equal-time quantization, particularly to study the bound state problems
or the strongly-coupled systems.

1) First of all, the fact that the {\it vacuum is trivial} with 
{\it no vacuum polarization} 
suggests a novel way of understanding the non-perturbative
phenomena such as the confinement and the spontaneous chiral symmetry breaking
which are due to complicated vacuum structure 
in the conventional picture of the equal-time quantization.  
Particularly, the absence
of vacuum polarization may naturally explain 
the success of {\it non-relativistic picture} of the constituent quark model.
 
2) Second, we may introduce the {\it wave function} \cite{Leut}
$\psi_P(\vec{p}_1,\vec{p}_2, \cdot\cdot\cdot, \vec{p}_n)$
for the relativistic bound states $\vert P\rangle$ which are to be
expanded in terms of the physical Fock space 
(\ref{Fock}):
\begin{equation}
 \vert P\rangle = \sum_{n} \int 
 \prod_i^n \left( \frac{d \vec{p}_i}{(2\pi)^3 2 p_i^+}\right)
\psi_P(\vec{p}_1,\vec{p}_2, \cdot\cdot\cdot, \vec{p}_n)
\delta(\vec{P}-\sum_{n}\vec{p}_n) \vert \vec{p}_1,\vec{p}_2, \cdot\cdot\cdot, \vec{p}_n\rangle.
\label{composite}
\end{equation}
 Actually,
the creation operators of quarks and gluons in QCD are nothing but the 
partons; The probability $f_k (x)=
\int\prod_i 
(d \vec{p}_i/(2\pi)^3 2 p_i^+)
 |\psi_P(\vec{p}_1,\vec{p}_2, 
\cdot\cdot\cdot, \vec{p}_n)|^2 \delta(x-x_k)$, 
with $x_i\equiv p^+_i/P^+$, is the parton 
distribution function.
 On the other hand, in the equal-time quantization the concept of wave
 function makes
 sense only in the nonrelativistic limit where the number of
 particles is conserved and the vacuum polarization is absent.
  
3) Third,  the triviality of the vacuum suggests that 
{\it even in the NG phase}  the LF charges
$Q_{\alpha}\equiv\int d \vec{x} J^+_{\alpha}(x), 
Q_{5\alpha}\equiv\int d \vec{x} J^+_{5\alpha}(x)
$
do {\it annihilate the vacuum}:
\begin{equation}
Q_{\alpha}\vert 0\rangle =
Q_{5\alpha}\vert 0\rangle =0,
\label{JSTh}
\end{equation}
where $J^{\mu}, J^{5\mu}$ are vector and axial-vector currents, respectively.
Namely, the vacuum is singlet under the transformation generated by these charges.
This fact was known for long time as Jers\'ak-Stern theorem 
which was proved when massless particle is absent (i.e., {\it not} 
in the chiral symmetric limit) \cite{JS,MY} \footnote{
~We shall later show in DLCQ that (\ref{JSTh}) actually 
holds even in the chiral symmetric
limit with massless NG boson \cite{KTY,TY}. 
}
.
Thus the chiral $SU(3)_L\times SU(3)_R$ algebra of LF charges,
\begin{eqnarray}
\left[ Q_{\alpha},Q_{\beta} \right] & = & 
i f_{\alpha\beta\gamma}Q_{\gamma}, \nonumber \\
\left[ Q_{\alpha}, Q_{5\beta} \right] & = & 
i f_{\alpha\beta\gamma}Q_{5\gamma}, \nonumber \\
\left[ Q_{5\alpha}, Q_{5\beta} \right] & = & 
i f_{\alpha\beta\gamma}Q_{\gamma},
\label{chiralalgebra}
\end{eqnarray}
with $\alpha,\beta,\gamma=1,2,\cdot \cdot \cdot, 8$,
may be useful to classify the physical states which are constructed
out of a chain of field operators acting on the vacuum as (\ref{Fock}). 
Since the vacuum is singlet and the field operators have definite 
transformation property of this algebra, the physical states also 
have definite transformation property. This is in contrast to
the equal-time quantization where the vacuum in NG phase is not singlet and
hence the physical states do not have definite transformation property even
though the field operators do. Actually the chiral algebra of LF charge
is nothing but the algebraization of the celebrated 
Adler-Weisberger (AW) sum rules \cite{AW} 
in hadron physics \cite{Ida,CHKT}. Through analysis of AW sum rules, 
hadrons are in fact classified into
({\it reducible}) representations of the LF chiral algebra,
which coincides with the one time fashionable  ``representation 
mixing'' \cite{GH,Weinberg,Ida,Melosh,CHKT}.
We shall  fully discuss this point
in Section 3.

\subsection{Zero-Mode Problem}
The above arguments, however, are subject to 
criticism that the zero mode with $p^+=0$ has been simply ignored. 
Actually, from the spectral condition (see Fig.1) there is 
{\it no gap} between the zero mode $p^+=0$
and the non-zero modes $p^+ >0$ and hence it is rather delicate whether 
the zero mode can be neglected or
not. If there exists a zero-mode state degenerate
with the trivial vacuum, then the trivial vacuum may not be
the true vacuum:
\begin{eqnarray}
P^+\vert 0\rangle =0 \not \Rightarrow \
P^-\vert 0\rangle=0,
\label{ntruev} 
\end{eqnarray}
  in contrast to (\ref{truev}).

Historically, it was first argued by Weinberg \cite{WIMF}
in the perturbation theory that 
the  ``vacuum diagrams'' (diagrams with lines created/annihilated 
from/into the vacuum and
with external lines, Fig.2)  and
the vacuum polarization (vacuum to 
vacuum diagram without external lines, Fig.3) 
both vanish in the infinite-momentum frame. 
The result was then re-examined through the Feynman rule 
in terms of LF parameterization \cite{CM};
\begin{equation}
\int d\mbox{\boldmath$p$}dp^0 \rightarrow
\int d\vec{p}dp^-,
\end{equation}
with the propagator
\begin{eqnarray}
\Delta_F(x;m^2)&=&\theta(x^+) \Delta^{(+)}(x;m^2)-
\theta(-x^+)\Delta^{(-)}(x;m^2)\nonumber \\
&=&\frac{1}{(2\pi)^4} \int d \vec{p} \int d p^-
\frac{e^{-ip^- x^+ +i\vec{p}\vec{x}} }{2p^+p^- - p_{\bot}^2 -m^2+i0}, 
\end{eqnarray}
where
\begin{equation}
\Delta^{(\pm)}(x;m^2)
=\pm \int \frac{d \vec{p} dp^-}{(2\pi)^3} 
\theta (\pm p^+)\delta(2p^+ p^-  - p_\bot^2 -m^2)
e^{-i(p^- x^+ + \vec{p}\vec{x})}.
\label{deltaplus} 
\end{equation}
Here  we note that $\pm \theta (\pm x^+)\Delta^{(\pm)}(x;m^2) 
\sim 
 \theta (\pm x^+)\theta(\pm p^+)
\sim \theta (x^+ p^+)$
apart form the factor $\frac{1}{(2\pi)^3}\int d^4 p\delta(p^2-m^2) e^{-ipx}$.
Note that $\theta(p^+ x^+)$ instead of $\theta (p^- x^+)$, in sharp
contrast to $\theta (p^0 x^0)$ in the equal-time quantization.
Then the particle lines with $x^+>0$ carry $p^+>0$, 
while the lines with $x^+<0$ which 
carry $p^+<0$ are interpreted as anti-particles with $x^+>0, p^+>0$. 
This actually led to vanishing 
vacuum diagrams with external lines (Fig.2),
since $P^+$ conservation (momentum conservation)
forbids the transition between the vacuum 
($p^+=0$) and the intermediate states ($p^+=\sum_{k} p_k^+>0$ in $x_k^+>0$ 
direction) with each internal line having $p_k^+>0$:  
\begin{equation}
({\rm Fig.2}) 
\quad =\quad 0.
\end{equation}

\begin{figure}[htbp]
  \begin{center}
    \leavevmode
\epsfxsize=3cm
    \epsfbox{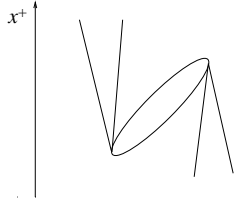}
  \end{center}
\caption{Vacuum graph with external lines.}
\label{fig:2}
\end{figure}
%
\begin{figure}[htbp]
  \begin{center}
    \leavevmode
\epsfxsize=2cm
    \epsfbox{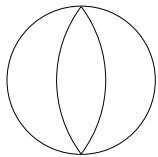}
  \end{center}
\caption{Vacuum polarization.}
\label{fig:3}
\end{figure}
On the contrary,
the vacuum polarization (graph {\it without external lines}), Fig.3, 
does {\it not} vanish due to the {\it zero-mode contribution} \cite{CM}:
\begin{eqnarray}
({\rm Fig.3})
&=&\int dp^+ \int d p^- \frac {F(p^+p^-)}{2p^+p^- -m^2 +i 0}\nonumber\\
&=&\int_{-\infty}^{\infty}d\lambda 
\tilde{F}(\lambda) \int_{0}^{\infty} d\xi \frac{e^ {-im^2\xi}}{i} \int dp^+
\int dp^-  e^{i 2p^+p^-(\xi+\lambda)}\nonumber \\
&=& C
\int d p^+ \delta(p^+)\quad  \ne\quad 0 ,
\label{nzerovp0}
\end{eqnarray}
where 
$F$ and $\tilde{F}$ are a certain function and its Fourier transform
with respect to $\lambda$, respectively and  
$C= -
\pi i \int d\lambda d\xi e^ {-im^2\xi} \tilde{F}(\lambda)/  
(\lambda +\xi) $
a numerical constant, and
we have disregarded the transverse part which is irrelevant.
 
Next we discuss the axiomatic arguments to guarantee the physical Fock space by
removing the zero mode through restricting to the localized wave packet 
without zero mode \cite{RS}:
\begin{equation}
\int_{-\infty}^{\infty} dx^- f_{\alpha}(\vec{x}) =0  
\quad (p^+> 0),
\label{testnozero}
\end{equation}
instead of the 
``plane wave'' $e^{-i(p^+x^- -p^{\bot}x^{\bot})}/(2\pi)^3 2p^+ 
$. 
Now $\phi(x)$ may be expanded as
\begin{equation}
\phi(x)=\sum_{\alpha}f_{\alpha}(\vec{x})a_{\alpha}(x^+)
+f^{*}_{\alpha}(\vec{x})a^\dagger_\alpha(x^+),
\end{equation}
where
\begin{equation}
a_\alpha (x^+) =i\int d\vec{x}f^{*}_{\alpha}(\vec{x})
 \stackrel{\leftrightarrow}{\partial}_- \phi(x)
\end{equation}
has been used in place of (\ref{invFT}).
Then the canonical commutator (\ref{CCR}) yields the algebra of
creation and annihilation operators, $a_{\alpha}$ and 
$a^\dagger_{\alpha}$:
\begin{equation}
\left[a_{\alpha},a^\dagger_{\beta}\right] = 
i\int d\vec{x}f^{*}_{\alpha}(\vec{x})
 \stackrel{\leftrightarrow}{\partial}_- f_\beta(\vec{x})= 
\delta_{\alpha\beta}.
\end{equation}
Since $a_{\alpha}$ has no zero mode, we expect it in fact  annihilates 
the vacuum:
\begin{equation}
a_{\alpha}(x^+)\vert 0\rangle=0
\end{equation}
without recourse to the dynamics 
and physical Fock space without zero mode 
is constructed upon this trivial vacuum.

However, this trick is actually not successful \cite{NY}, since 
it would lead to the two-point Wightman function on LF as
\begin{equation}
\langle 0\vert \phi(x) \phi(0)\vert0|\rangle|_{x^+=0}
=
\sum_{\alpha} f_{\alpha}(\vec{x}) f^\ast_{\alpha} (\vec{0}),
\label{W-fnozero}
\end{equation}
which does not hold as seen from below.
When integrated over $x^-$, the R.H.S. of (\ref{W-fnozero}) yields 0
(see (\ref{testnozero})),
\begin{equation}
\sum_\alpha \int_{-\infty}^{\infty} dx^- f_\alpha(\vec{x})f_\alpha^\ast(\vec{0})=0.
\end{equation}
On the other hand, in the free theory
the two-point function on L.H.S. is explicitly given by $\Delta^{(+)}(x;m^2)$ 
in (\ref{deltaplus}) which is written in terms of the Hankel
function $K_1$ in the space-like region $x^2 <0$: 
\begin{eqnarray}
\langle 0|\phi(x)\phi(0)|0 \rangle  
&=&\Delta^{(+)}(x;m^2)
\nonumber\\ 
&=& \frac{1}{(2\pi)^3}\int_{0}^{\infty}
\frac{dp^+ }{2p^+}\int_{-\infty}^{\infty} dp^{\perp} 
 e^{-i(
\frac{m^2+p_\bot^2}{2p^+}x^+
+\vec{p}\vec{x})}
\nonumber\\
&=&\frac{m}{4\pi^2\sqrt{-x^2}}K_1 (m\sqrt{-x^2}) \quad\quad
(x^2<0).
\label{WFcov}
\end{eqnarray}
Restricting (\ref{WFcov}) to the LF, $x^+=0$, yields 
\begin{equation}
\left. \langle 0|\phi(x)\phi(0)|0\rangle \right|_{x^+=0}
=\frac{m}{4\pi^2\sqrt{x_{\bot}^2}}K_1 (m\sqrt{x_{\bot}^2}),
\label{WFcov2}
\end{equation}
which is {\it $x^-$-independent}. Then L.H.S. of (\ref{W-fnozero}) 
integrated over $x^-$ does not
vanish 
in the space-like region on LF ($x^2=-x_{\bot}^2<0$):
\begin{equation}
\int_{-\infty}^\infty  dx^- \Delta^{(+)}(\vec{x};m^2)
=\int_{-\infty}^\infty  dx^-\frac{1}{4\pi^2}\frac{m}{\sqrt{x_{\bot}^2}}K_1(m\sqrt{x_{\bot}^2})\ne 0,
\end{equation}
which is again the zero-mode contribution $\sim \delta(p^+)\vert_{p^+=0}$
(see (\ref{nzerovp0})).
Then (\ref{W-fnozero}) is in self-contradiction. Namely,
$\{f_{\alpha}\}$ is not a complete set.  

Thus these two examples strongly suggest that {\it we cannot neglect the
zero mode in a way consistent with the  Lorentz invariance}.
We shall later demonstrate that it is indeed the case (no-go theorem 
\cite{NY}).  
Moreover, the zero mode is {\it a necessity} for the NG phase in the LF
quantization: In the sigma model, for example,  the NG phase 
takes place when the scalar field $\phi(x)$
develops the vacuum expectation value $\langle \phi(x) \rangle$,
which is independent of $x^{\mu}$ and hence is nothing but the zero mode
contribution. Thus the {\it zero mode is the only mode responsible for the 
NG phase}: We should not ignore it when discussing the NG phase. So simply ignoring the
zero mode will lose important physics besides losing consistency
with the Lorentz invariance.
If there existed such a zero mode state, on the other hand,
it would be degenerate with the vacuum with respect to the 
eigenvalue of $P^+$ and hence the 
vacuum would not be 
identified through the $P^+$ spectrum until 
we solve $P^-$ literally, namely (\ref{ntruev}). This however implies giving up
the physical Fock space, the best advantage of the LF quantization.
Thus the zero mode would be the most dangerous obstacle against the LF approach.

Then, what happens to the trivial vacuum and the physical Fock space,
if one {\it takes account of the zero mode instead of simply ignoring it}?
This is the ``zero-mode problem'' which was first addressed by MY\cite{MY}.
The problem in the continuum theory
just discussed is that there is no gap between the zero mode and non-zero modes,
and hence the zero mode cannot be treated in a well-defined manner.
This problem was solved in DLCQ\cite{MY} which is the subject of the next section.

%% file: dpnu972.tex
\section{Discrete Light-Cone Quantization (DLCQ)}
\my
The Discrete Light-Cone Quantization (DLCQ) was proposed
by  Maskawa and Yamawaki (MY)\cite{MY} 
 to solve the above zero-mode problem without ambiguity:
 $x^-$ is confined into a finite region $-L\le x^-\le L$ with
a periodic B.C., and hence
the $P^+$ spectrum becomes {\it discrete}: 
\begin{equation}
p^+=n\frac{\pi}{L}\quad (n=0,1,2,\cdot \cdot \cdot).
\end{equation}
Then the zero mode $(n=0)$ is 
treated in a clean separation (gap) from other
modes ($n \ne 0)$. 
As it turns out shortly, the LF 
quantization is necessarily the quantization of the constrained system
whose canonical quantization is done through the Dirac's
quantization for the constrained system\cite{Dirac2} (For the 
LF quantization \`a la Dirac, see reviews \cite{HRT}).
Then MY  formulated the DLCQ via Dirac's quantization and  
discovered  the {\it zero mode constraint}, as an extra (secondary) second-class
constraint for the zero mode, by which the zero mode 
becomes an auxiliary field and can be eliminated out of the Fock space, and 
thereby 
{\it established the trivial vacuum and 
the physical Fock space} in DLCQ without recourse to the dynamics. 
The ``continuum" limit $L \rightarrow \infty$ 
(or, more precisely, infinite volume limit) is 
taken {\it only at the final 
stage of the whole calculations}. 
DLCQ was also considered independently by Casher\cite{casher}
in a somewhat different context (without Dirac's quantization and ignoring the
zero mode). 
It was later advocated by Pauli and
Brodsky\cite{PB} in solving the bound state problem (without attention to
the zero mode).
Here we depict the MY paper\cite{MY} in some detail, which is the basis of
the whole DLCQ physics.

\subsection{DLCQ 
- Canonical Formalism}
Let us consider the self-interacting scalar 
theory in $(3+1)$ dimensions 
whose Lagrangian is 
expressed in terms of the LF coordinate as  
\begin{equation}
{\cal L}={\partial}_{+} \phi {\partial}_{-} \phi 
-\frac{1}{2}({\partial}_{\bot}\phi)^2-\frac{1}{2}
{\mu}^2 {\phi}^2-V(\phi), 
\label{sl}
\end{equation}
where $V(\phi)$ is a potential. 
The Euler-Lagrange equation reads
\begin{eqnarray}
\left( 2\partial_+\partial_- -
\partial_\bot^2
 +\mu^2\right)\phi(x)
= - \frac{\partial V}{\partial \phi},
\label{ELeq}
\end{eqnarray}
which is the {\it first-order} differential equation with respect to
the LF time $x^+$ in contrast to the second-order $\partial_0^2$ 
in the equal-time
formalism. 
Accordingly, the canonical momentum conjugate to $\phi(x)$, 
\begin{equation}
\pi_\phi (x)=\frac{\partial {\cal L}}{\partial({\partial}_+ \phi)}
=\partial_{-}\phi(x),  
\label{primary0}
\end{equation}
is not an independent variable, since $\partial_-$ is
not a time-derivative.
Then this is a {\it constrained system}\footnote{\samepage{%
~This should {\it not} be confused with reduction of the
physical degree of freedom: The information of $\phi(x)$ on LF
just corresponds to that of a set of $(\phi(x), \pi_\phi (x))$ with
$\pi_\phi (x) \equiv \partial_0\phi(x)$,
in the equal-time
formalism. In the free theory, for example,
$(a, a^\dagger)$ in the equal-time quantization
are defined only through a set of $(\phi(x), \pi_\phi(x))$, while in the LF
quantization they are
defined by $(p_n^+>0, p^+<0)$ modes of $\phi(x)$ alone as in
(\ref{FT}).}}.
The canonical quantization of such a system is in fact done through the
Dirac's method for the constrained system.
Eq. (\ref{primary}) yields a primary constraint 
of the theory:
\begin{equation}
\Phi(x)=\pi_\phi (x)-\partial_{-}\phi(x)\approx 0.  
\label{primary}
\end{equation}

Since $x^-$ is restricted to the finite region $-L\le x^- \le L$, 
the 
B.C. should be specified at 
$x^{-}=\pm L$. We adopt the {\it periodic B.C.
} 
on $x^-$, which is consistent with the {\it $P^+$ conservation} and 
the {\it non-vanishing vacuum expectation value} of the scalar field
to be require by the NG phase. 
In fact, very existence of the zero mode is related to 
this periodic B.C..
Then we write the Fourier expansion:
\begin{eqnarray}
\phi(x)&=&\varphi(x)+\phi_0(x^+,x^{\bot}), 
\label{dec-1} \\
\varphi (x)&=&\frac{1}{2L}\sum_{n > 0}\frac{1}{2p_n^+}
\int \frac{d p^\bot}{(2\pi)^2}
\left[ a_n(p^{\bot},x^+) e^{-i (p_n^+ x^- - p^\bot x^\bot)
}
\right.
\nonumber \\
& & 
\left. +a _n^{ \dagger}(p^{\bot},x^+) e^{i(
(p_n^+ x^- - p^\bot x^\bot)
} \right],
\label{exp} \\
\phi_0 (x^+,x^{\bot})&\equiv& \frac{ 1}{2L}\int_{-L}^{L}\phi(x)dx^{-},
\label{zero-mode} 
\end{eqnarray}
with $p_n^+=n\pi/L \quad (n=1,2,\cdot \cdot \cdot)$,
which is similar to the Fourier transform (\ref{FT}) except for
the explicit separation of the zero mode $\phi_0
(x^+,x^{\bot})$ from 
the oscillating modes $\varphi (x)$. 

Here we follow a slightly different formulation\cite{TY}
from the original one\cite{MY}
based on an explicit orthogonal decomposition\cite{H} of 
the primary constraint into two parts, zero mode and non-zero mode.
According to the decomposition of the scalar field $\phi(x)$ into  
the oscillating modes $\varphi (x)$ plus the zero mode $\phi_0
(x^+,x^{\bot})$,
the conjugate momentum 
$\pi_\phi$ may also be divided as 
\begin{equation}
\pi_\phi (x)=\pi_{\varphi}(x)+\pi_0 (x^+,x^{\bot}),  
\label{dec-2}
\end{equation}
where $\pi_0$ and $\pi_{\varphi}$ are the zero modes conjugate to 
$\phi_0$ and that to the remaining orthogonal part $\varphi(x)$,
respectively. 
Now, substituting (\ref{dec-1}) and (\ref{dec-2}) into 
(\ref{primary}), we have two independent 
constraints, 
\begin{equation}
\Phi_{1}(x)\equiv \pi_{\varphi}(x)-\partial_{-}
\varphi(x)\approx 0
\label{Phi1} 
\end{equation}
and 
\begin{equation}
\Phi_{2}(x)
\equiv \pi_{0}(x^+,x^{\bot})\approx 0,  
\label{Phi2}
\end{equation}
in place of the original one (\ref{primary}).

Then the the fundamental Poisson bracket, 
\begin{equation}
\{ \phi (x),\pi_\phi (y)\}
=\sum_n \frac{1}{2L} e^{i\frac{n\pi}{L} (x^- - y^-)}\cdot
\delta
(x^\bot-y^\bot)
= \delta
(\vec x-\vec y) , 
\label{poisson1}
\end{equation}
may also be divided into the non-zero mode and the zero mode:    
\begin{eqnarray}
\{\varphi(x),\pi_{\varphi}(y)\}&=&
\sum_{n\ne 0} \frac{1}{2L} e^{i\frac{n\pi}{L} (x^- - y^-)}\cdot
\delta
(x^{\bot}-y^{\bot})\nonumber\\
&=&\left[\delta(x^-
-y^-)-\frac1{2L}\right]
\delta
(x^{\bot}-y^{\bot}),
\label{poisson3} \\
\{\phi_0,\pi_0\}&=&\frac{1}{2L}\delta
(x^{\bot}-y^{\bot}), 
\label{poisson2}
\end{eqnarray}
where $x^+=y^+$ is understood.
All other Poisson brackets are equal to zero as expected.   
Thus the Poisson brackets between the constraints are
\begin{eqnarray}
\{\Phi_1(x),\Phi_1(y)\} &=& (\partial_-^x - \partial_-^y)\delta(x^- - y^-)
   \cdot\delta(x^\bot -y^\bot),
\label{PB11}   \\
\{\Phi_1(x),\Phi_2(y)\}&=&\{\Phi_2(x),\Phi_1(y)\}=\{\Phi_2(x),
\Phi_2(y)
\}=0. 
\label{PB12}
\end{eqnarray}

The total Hamiltonian is obtained by adding the primary constraints
to the canonical one $H_c$: 
\begin{eqnarray}
H_T &\equiv& H_c+\int d \vec{x}\hspace{5pt} 
\left[ v_1(x)\Phi_1(x)+
v_2(x) \Phi_2(x) \right] ,\\
H_c&=& \int d \vec{x} \hspace{5pt}
\left[\frac{1}{2}\{(\partial_{\bot}\phi)^2 
+\mu^2 \phi^2\}+V(\phi)\right], 
\end{eqnarray}
where $v_2$ and $v_1$ are the zero mode and 
the remaining part of the Lagrange multiplier, respectively. 
The multiplier $v_1$ (without zero mode)
is determined by the consistency condition 
for $\Phi_1(x)$:
\begin{equation}
\dot{\Phi}_1(x)=\{\Phi_1(x),H\}=(\partial_\bot^2 - \mu^2) \phi(x)
-\frac{\partial V}{\partial \phi} - 2\partial_-v_1(x) \approx 0,
\end{equation}
which can be easily integrated without ambiguity owing to the periodic
B.C..
On the other hand, the consistency condition for 
$\Phi_2(x)$, 
\begin{equation}
\dot {\Phi}_2(x)=\{\Phi_2(x),H_T\}
={1 \over 2L}\int_{-L}^L dx^{-}
\left[(\partial_{\bot}^2-\mu^2)\phi-
{\partial V \over \partial \phi}\right] \approx 0,
\end{equation}
leads to a new constraint so-called 
``{\it zero-mode constraint}" \cite{MY}: 
\begin{equation}
\Phi_3 (x) \equiv \frac{1}{2L}\int_{-L}^L dx^{-}
\left[(\mu^2-\partial_{\bot}^2)\phi+\frac{\partial V}{\partial
\phi}\right] \approx 0.  
\label{zmconstraint}
\end{equation}
The consistency condition for the zero-mode constraint 
yields no further constraint and just 
determines the multiplier $v_2$. 
Note that in deriving these relations we have used the condition
\begin{equation}
\delta(x^--L)=\delta(x^-+L), 
\end{equation}
which comes from the definition of the delta function with the periodic
B.C.:
\begin{equation}
\delta(x^-)=\frac{1}{2L}\sum_{n \in {\bf Z}} e^{\frac{in\pi}{L}x^-}.  
\end{equation}

Having obtained all the second-class 
constraints $\Phi_1$ ((\ref{Phi1})), $\Phi_2$ ((\ref{Phi2})) and 
$\Phi_3$ ((\ref{zmconstraint})), we are ready to calculate the  
Dirac bracket. The Dirac bracket is the Poisson bracket calculated in terms of a
reduced set of unconstrained canonical variables, with the redundant 
degree of freedoms being subtracted through the second class 
constraints \cite{MN},
and hence is the basis of the canonical quantization for the constrained
system via the correspondence principle between the Dirac bracket and
the commutator:
\begin{eqnarray}
i\hbar \{A, B \}_D \rightarrow [A, B].
\end{eqnarray}
The Dirac bracket of two arbitrary 
dynamical variables $A(x)$ and $B(y)$ is given as 
\begin{eqnarray}
&&
\{A(x),B(y)\}_{D} \equiv \{A(x),B(y)\} 
                  -\sum_{i,j=1}^3 \int d \vec{u} 
                  d \vec{v} \nonumber \\
&&            \cdot       \{A(x),\Phi_i(u)\}
                     (C^{-1})_{i,j}(u,v)\{\Phi_j(v),B(y)\},  
\label{DM}
\end{eqnarray}
where $ (C^{-1})_{i,j} (x,y)=-
(C^{-1})_{j,i}(y,x)$ is the inverse of 
$C_{i,j}(x,y)\equiv \{\Phi_i(x),\Phi_j(y)\}=-C_{j,i}(y,x)$ which is
the matrix of Poisson brackets of the constraints:
\begin{equation}
C_{i,j}(x,y)=\left(
\begin{array}{ccc}
a(\vec{x},\vec{y}) & 0 & b(\vec{x},y^\bot)\\
0                  & 0 & c(x^\bot,y^\bot)\\
-b(\vec{y},x^\bot)&- c(y^\bot,x^\bot)&0
\end{array}
\right),
\end{equation}
where
\begin{eqnarray}
a(\vec{x},\vec{y})&=&
(\partial^{y}_{-}-\partial^{x}_{-})\delta
(\vec{x}-\vec{y})
\nonumber \\
&=&\frac{1}{2L}\sum_{n\in {\bf Z}}\left(\frac{-2in \pi}{L}\right)
\hspace{5pt}e^{\frac{in\pi}{L}
(x^- -y^-)}
\cdot\delta
(x^{\bot}-y^{\bot}),
\nonumber\\
b(\vec{x},y^\bot)&=&
-\frac{1}{2L}\left( \alpha(\vec{x})-\frac{1}{2L}\beta^{-1}(x^\bot)\right)
\delta
(x^{\bot}-y^{\bot}),
\nonumber\\
c(x^\bot,y^\bot)&=&
-\frac{1}{4L^2}\beta^{-1}(x^\bot)
\delta
(x^{\bot}-y^{\bot}),
\end{eqnarray}
with
\begin{equation}
\alpha(\vec{x})
\equiv\mu^2-\partial_{\bot}^2+\frac{\partial^2 V}{\partial \phi^2}, 
\quad \quad 
\beta^{-1}(x^{\bot})
\equiv\int^L_{-L} dx^{-}
\alpha(\vec{x}). 
\label{alphabeta}
\end{equation} 
Note that $\frac{1}{2L}\beta^{-1}$ is the
zero mode of $\alpha$, and 
$\Phi_3$ and $\Phi_2$ are a ``conjugate pair'' of
the second-class constraints for the zero mode.
In view of the Dirac bracket, $\{A(x),\Phi_i(y)\}_D=0$ for the 
arbitrary dynamical variable $A(x)$, so that the constraints become
strong relations; $\Phi_i(x)\equiv 0$ (strongly).
 
Now we calculate the 
inverse matrix $C^{-1}_{i,j}(x,y)$. Straightforward calculation 
 yields \cite{TY}:
\begin{equation}
C^{-1}_{i,j}(x,y)=\left(
\begin{array}{ccc}
a^{-1}(\vec{x},\vec{y}) & p(\vec{x},y^{\bot}) & 0\\
-p(\vec{y},x^{\bot})    & 0                   & -c^{-1}(y^\bot,x^\bot)\\
0                       &c^{-1}(x^\bot,y^\bot)&0
\end{array}
\right),
\end{equation}
with 
\begin{eqnarray}
a^{-1}(\vec{x},\vec{y})&=& -\frac{1}{4}\left( \epsilon(x^- - y^-) 
- \frac{x^- - y^-}{L}\right)\cdot \delta
(x^\bot -y^\bot)\nonumber\\
&=&\frac{1}{2L}
\sum_{n \ne 0, n \in {\bf
Z}}
\left(\frac{-L}{2in\pi}\right)\hspace{5pt}
e^{\frac{in\pi}{L}(x^--y^-)}\nonumber\\
& &\cdot\delta
(x^{\bot}-y^{\bot}),
\label{ainverse}
\\
p(\vec{x},y^\bot)&=& -\int d\vec{u} d v^\bot a^{-1}(\vec{x},\vec{u})
b(\vec{u},v^\bot)
c^{-1}(v^\bot,y^\bot)\nonumber\\
&=&\frac{L}{2}\int_{-L}^{L}d u^-
\left[\epsilon(x^- -  u^-)-\frac{x^- - u^-}{L}\right]\nonumber\\
& &\cdot \beta(y^\bot) \alpha(u^-,y^\bot)\delta
(x^\bot-y^\bot), \\
c^{-1}(x^\bot,y^\bot)&=& -4L^2 \beta(x^\bot) \delta
(x^\bot -y^\bot),
\end{eqnarray}
where we noted
\begin{equation}
\int_{-L}^{L} du^-  \left[\epsilon(u^--v^-)- \frac{u^--v^-}{L}\right] 
=0      
\label{intsign}
\end{equation}
and $0$'s in the diagonal entry can be read off from the one-dimensionality
in the $x^-$ ($y^-$)-space of the anti-symmetric matrix.
Note that 
$a^{-1}$ is the inverse of $a$ in the sense that
\begin{eqnarray}
\int
d\vec{z}
a (\vec{x},\vec{z}) a^{-1}(\vec{z},\vec{y})
&=&
\frac{1}{2L}\sum_{n \ne 0, n \in {\bf Z}}e^{i\frac{n\pi} 
                {L}(x^{-}-y^-)} 
\cdot\delta
(x^\bot -y^\bot)\nonumber\\
&=&
\left(\delta (x^- - y^-) -\frac{1}{2L}\right)
\cdot\delta
(x^\bot -y^\bot),
\end{eqnarray}
where $-1/2L$,
as well as $-(x^- - y^-)/L$ in (\ref{ainverse}),
stands for (the subtraction of) the zero-mode contribution
(Note also that $\partial_-^x \epsilon (x^- -y^-) = 2\delta(x^- -y^-)$). 

Recalling  (\ref{DM}) and  (\ref{poisson3}) - (\ref{poisson2}), 
we obtain the Dirac bracket $\{A,B\}_{D}$ and hence the commutator
$[A,B]=i\{A,B\}_{D}$ apart from the operator ordering \cite{MY,TY}:
\begin{eqnarray}
[\varphi(\vec{x}),\varphi(\vec{y})]
&=&i\, a^{-1}(\vec{x},\vec{y})
\nonumber\\
&=&
-{i \over 4}\left[\epsilon(x^--y^-)-{x^--y^- 
\over L}\right]
\delta
(x^{\bot}-y^{\bot})
\label{varphicomm}
\end{eqnarray}
and
\begin{eqnarray}
 {}[\varphi(\vec{x}), \phi_0(y^{\bot})]
&=& \frac{i}{2L}\cdot
p(\vec{x}, y^{\bot})\nonumber \\
&=&\frac{i}{4}\int_{-L}^{L}du^-
\left[\epsilon(x^--u^-)-{x^--u^- \over L}\right]
\nonumber \\
&&\cdot \beta(y^{\bot}) \alpha(u^-,y^{\bot}) 
\delta
(x^{\bot}-y^{\bot}),
\label{var-0comm}\\
{}[\phi_0(x^\perp),\varphi(\vec{y})]
&=& 
{}-[\varphi(\vec{y}),\phi_0(x^{\bot})],
\label{0-varcomm}\\
{} [\phi_0(x^\bot), \phi_0(y^\bot)]
&=&0.
\label{0-0comm}
\end{eqnarray}
Combining these, we arrive at the canonical DLCQ commutator\cite{MY} for
the full field $\phi =\varphi +\phi_0$ up to operator 
ordering:
\begin{equation}
\left[\phi(x),\phi(y)\right]
=-\frac{i}{4}\left[\epsilon(x^--y^-)-2\beta \int_{y^-}^{x^-}
                   \alpha(z^-)dz^- \right]\delta
                     (x^{\bot}-y^{\bot}).
\label{comm2}
\end{equation}

Eq.(\ref{varphicomm}) coincides with
the commutator of the full field 
in the free theory 
\cite{MY}, since the
$x^-$-dependence of $\alpha(\vec{x})$ drops out and hence 
the integral in (\ref{var-0comm}) yields zero contribution because of
(\ref{intsign}):
\begin{eqnarray}
[\phi(x),\phi(y)] =-\frac{i}{4}\left[
\epsilon (x^- - y^-) -\frac{x^- -y^-}{L}\right]
\delta(x^\perp - y^\perp).
\label{freecomm}
\end{eqnarray}
As to the commutator of the full field 
(\ref{comm2}), it  is {\it not a c-number but an 
operator}, 
since $\alpha$ and $\beta$ given in (\ref{alphabeta}) are operators. This
peculiarity is due to the zero mode whose commutator with the non-zero mode
is actually the operator as seen in 
(\ref{var-0comm}), (\ref{0-varcomm}), in sharp contrast to the {\it 
commutator of 
the oscillating modes} (\ref{varphicomm})
 which is a c-number. 

Note that, compared with the standard commutator (\ref{CCR}) 
in the continuum theory, we have an extra term $-(x^- - y^-)/L$ in 
(\ref{varphicomm}) and (\ref{freecomm}),
which of course stands for the subtraction of the zero mode.
This extra term as it stands is formally of order $O(1/L)$. Then one
might consider that such a term could be neglected
in the continuum limit $L \rightarrow \infty$ and the DLCQ commutator 
(\ref{varphicomm}) would coincide with the standard one
(\ref{CCR}). However, it
is not true, since the commutator is usually used in the {\it quantity 
integrated over }
$\int_{-L}^{L} d x^- $ like LF charge and Poincar\'e generators, in which
case the contribution of
the extra term becomes of order $O(1)$ after such an integration, as we shall
discuss later. Here we present just a typical example: Eq.(\ref{freecomm})
yields
\begin{eqnarray}
\int d\vec{x} \left[\partial_- \phi(x), \phi(y)\right]&=& 
\int_{-L}^L d x^- 
\frac{-i}{4}\left[2\delta(x^- - y^-) -\frac{1}{L} \right] \nonumber \\
&=&\frac{-i}{4} (2-2)=0
\label{polecr}
\end{eqnarray}
{\it independently of} $L$, whereas (\ref{CCR}) gives $-i/2 \ne 0$.
 As we stressed before, in DLCQ  
the continuum limit $L \rightarrow \infty$ must be taken 
{\it after whole calculations}. The same comment also applies to the
extra term in (\ref{comm2}) where $\beta=O(1/L)$.

\subsection{Zero-Mode Constraint and Physical Fock Space}

All these features actually reflect the {\it zero-mode constraint} $\Phi_3(x)$
((\ref{zmconstraint})) which is now a strong relation and hence
an operator relation as well \cite{MY}:
\begin{equation}
\frac{1}{2L}\int_{-L}^L dx^{-}
\left[(\mu^2-\partial_{\bot}^2)\phi+\frac{\partial V}{\partial
\phi}\right]  =0.
\label{zmconstraint2}
\end{equation}
{\it In the free theory, this zero-mode constraint dictates 
that the zero mode should vanish} identically, since $\mu^2-\partial_{\bot}^2$
is positive-definite:
\begin{eqnarray}
\phi_0 (x^+,x^\bot) =
\frac{1}{2L}\int_{-L}^L dx^{-}\phi(x)
=0
 \quad ({\rm free} \quad {\rm theory}),
\end{eqnarray}
in accordance with the above c-number commutator (\ref{freecomm}).
In the interacting theory, on the other hand,
the zero-mode constraint (\ref{zmconstraint2})
implies that the zero mode  
is {\it not an independent degree of freedom} but is
implicitly written in terms of non-zero (oscillating) modes
through interaction. 

It was actually the central issue of MY\cite{MY}
to argue that such a constrained zero mode can in principle be
{\it solved away out of the physical Fock space}  
and hence the {\it trivial LF vacuum is justified in DLCQ}. 
The 
commutator between
creation and annihilation operators 
$a_n^\dagger(p^\perp,x^+), a_n(p^\perp,x^+)$ 
of the non-zero modes 
are given by the Fourier transform of
(\ref{varphicomm}) via inverse Fourier transform of (\ref{exp})
similarly to (\ref{hocr}):
\begin{eqnarray}
[a_m (p^\perp, x^+), a_n^\dagger (q^\perp,x^+)]
=(2\pi)^2 2L\cdot 2p_n^+ \delta_{m, n} \delta(p^\perp -q^\perp).
\end{eqnarray}
Note again that these creation and annihilation operators are interacting
Heisenberg fields whose dynamics has not been solved.
Because of the periodic 
B.C.,
we have  translation
invariance $[\phi(x),P^+]=i\partial_- \phi(x)$, so that we have 
$[a_n (p^\perp,x^+),P^+]=p_n^+ a_n (p^\perp,x^+)$,
$[a_n^\dagger (p^\perp,x^+),P^+]=-p_n^+ a_n^\dagger (p^\perp,x^+)$
and hence
\begin{eqnarray}
a_n (p^\perp,x^+)\vert 0\rangle=0,  
\label{trivacDLCQ}
\end{eqnarray}
with $P^+\vert 0\rangle=0$, 
similarly to
(\ref{cr-ann})-(\ref{trivac}), except for the point that 
this time there is {\it no subtlety about the zero mode}.
Now, the zero mode  is in principle written in terms of the non-zero
modes through the zero-mode constraint (\ref{zmconstraint2}),
and hence we may write the Hamiltonian $P^-$ into the form:
\begin{eqnarray}
P^-  
= : F(a_m^\dagger, a_n) :,
\label{normalorder}
\end{eqnarray}
with $\sum_m p_m^+ -\sum_n p_n^+=0$ ($P^+$-conservation),
where $: \quad:$ stands for the {\it normal product} with respect
to the sign of $P^+$ instead of the energy $P^-$ and is {\it independent of the
dynamics}. The point is that due to the $P^+$-conservation,
$P^-$ should contain {\it at least one annihilation operator} which is to be
placed to the {\it rightmost} in the normal product: Namely,
\begin{eqnarray}
P^- \vert 0\rangle=0,
\end{eqnarray}
thus establishing that the trivial vacuum (\ref{trivacDLCQ})
in fact becomes the true vacuum \cite{MY}.

It is also noted \cite{MR} that the zero-mode constraint
(\ref{zmconstraint2})
can also be obtained by simply
integrating over $x^-$ the Euler-Lagrange equation (\ref{ELeq}) 
with use of the periodic 
B.C.:
\begin{equation}
0=-\int^{L}_{-L}dx^{-}2\partial_{+}\partial_{-}\phi
 =\int^{L}_{-L}dx^{-}\left[(\mu^2 -\partial_{\bot}^2)\phi +
 \frac{\partial V}{\partial \phi}\right] .
\label{zmconstraint3}
\end{equation} 
Namely, the zero mode constraint is {\it a part of the equation of motion} and
the zero mode is nothing but an auxiliary field having no kinetic term.

\subsection{Boundary Conditions}
We have imposed periodic B.C.,
$\phi (x^-=L)=\phi(x^-=-L)$,
in the finite box
$-L\leq x^- \leq L$. 
Here we discuss in general \cite{TY} 
the B.C.
of the LF quantization, {\it not only
the DLCQ but also the continuum theory}, which plays 
an essentially different role than that of the equal-time quantization.
Actually, as was emphasized by Steinhardt\cite{STH}
in the continuum  theory, 
the {\it B.C.
should always be specified, whether in the 
continuum theory or DLCQ}, 
in order to have a consistent LF quantization.
In fact, the B.C.
on LF includes a part of the dynamics 
in sharp contrast to 
the equal-time quantization. That is, {\it different 
B.C.
defines a different LF
theory}. For example, the periodic B.C.
allows scalar field to
develop the vacuum expectation value, while the anti-periodic
one does not, for the same Lagrangian.

Let us consider a scalar model 
{\it without B.C.
}
in either the ``continuum'' or 
``discrete'' LF quantization   
in the context of 
the Dirac quantization in Section 2. 
Without B.C.,
 the constraint for 
zero mode  will not appear. 
The only constraint appearing in the theory is (\ref{primary}),
$
\Phi(x)=\pi_\phi (x)-\partial_{-}\phi(x), 
$
whose Poisson bracket is given by  
$
\{\Phi(x), \Phi(y)\}=(\partial_{-}^y
-\partial_{-}^x)\delta (\vec{x}-\vec{y}), 
$
which looks the same as (\ref{PB11}) except for 
the non-separation of the zero mode.
Strictly speaking, we have infinitely many constraints which are  
expressible as linear combination of (\ref{primary}).

An important observation\cite{STH} is 
that there is a subset of constraint which might 
appear to be not only the first class but
also the second one.
To see this, consider a linear combination of 
the primary constraint (\ref{primary}):
\begin{equation}
\Phi_0 \equiv \int dx^- \Phi(x) , 
\end{equation}
which corresponds to the ``zero mode'' of $\Phi(x)$ 
in DLCQ. 
Suppose that any surface term is neglected throughout the calculation, 
then one  can  easily find   
\begin{equation}
\{\Phi_0, \Phi(x)\}=0 .
\label{1stcl} 
\end{equation}
This would mean that $\Phi_0$ might be first class, because 
it should commute with any linear combination of $\Phi(x)$ 
as a consistency. 
However, this is not always the case, 
as illustrated by the following example:  
\begin{eqnarray}
\{\Phi_0, \int \epsilon(y^-)\Phi(y)dy^-\}&=&-2\int dx^- dy^-
\partial_{-}^y 
 \epsilon(y^-)\delta(\vec{x}-\vec{y})
\nonumber\\
 &=&-4 
 \delta(x^{\bot}-y^{\bot})\ne 0   ,
\end{eqnarray}
where $\epsilon(x)$ is 
the sign function (\ref{sign}). This would mean that $\Phi_0$ 
might be second 
class in contradiction with the previous result (\ref{1stcl}). 
Actually,
  $\Phi_0$ is {\it neither 
first class nor 
second class}, which   simply
represents {\it inconsistency} hidden in the theory.   
This ambiguity manifests itself as the ambiguity 
of the inverse matrix of constraints, 
$C^{-1}$ in (\ref{DM}), and that of the Lagrange multiplier 
$v(x)$. It is easily shown that 
all such ambiguities can be removed,  
once the B.C.
at $x^-=\pm \infty$ or 
$x^-=\pm L$ is specified. 

Let us then study the 
allowed B.C.'s in DLCQ \cite{TY}. 
The same problem was studied by Steinhardt\cite{STH} 
within 
the continuum framework by neglecting all surface terms in
the 
partial integrations. Here  we study the same problem by 
carefully treating surface terms in DLCQ. 
For this purpose,  we  generalize $\Phi_0$ and consider the 
following constraint which appears in the total Hamiltonian: 
\begin{equation}
\Phi[v]=\int_{-L}^{L}dx^-v(x)\Phi(x) ,
\label{pc2}
\end{equation}
where $v(x)$ is a certain function (Lagrange multiplier) 
which satisfies the same B.C.
as $\phi(x)$ \cite{MY}. 
Once the B.C.
is specified,  
providing $\Phi[v]$ for all $v $ becomes  equivalent to 
providing $\Phi(x)$ for all $x$, which is nothing but the necessary
condition 
for consistency mentioned above. 
Moreover, we demand that 
the variation of canonical variable generated by (\ref{pc2}) 
must satisfy the same B.C..
We can derive this condition by writing down the functional 
variation of $\Phi[v]$:  
\begin{eqnarray}
\delta \Phi[v] 
= \int_{-L}^L d x^- [v(x) \delta 
\pi(x)+{\partial}_{-}v(x) \delta \phi(x)] \nonumber\\   
-v(x^-=L)\delta \phi(x^-=L)+v(x^-=-L)\delta \phi(x^-=-L) , 
\end{eqnarray}
where the first two terms on the R.H.S.  give the canonical variation 
of the fields which preserve the same B.C. 
as the canonical variables.  
On the other hands,  the surface terms generally violate the B.C..
One can thus require the condition  
\begin{equation}
v(x^-=L)\delta\phi(x^-=L)=v(x^-=-L)\delta\phi(x^-=-L) ,   
\label{gc}
\end{equation}
which is nothing but the discretized version of that derived
in the continuum theory\cite{STH}. 
This of course includes the periodic B.C.
we have studied.

Based on this condition  we investigate 
what kind of B.C.
can exist consistently.\cite{TY} 
We list up here some typical ones other than the periodic B.C.:
(I) the first boundary value: $\phi(x^-=L)=\phi(x^-=-L)=0$,\par
(II) the second boundary value:  
$\frac{d}{dx^-}\phi(x^-=L)=\frac{d}{dx^-}\phi(x^-=-L)=0$,\par
(III) the third boundary value: the mixed type of the above two 
conditions, \par
(IV) the anti-periodic B.C.,\\ 
where the right hand sides of both (I) and (II) can be generalized to 
any value. Note that $\delta\phi(x)$ and $v(x)$ obey 
the same B.C.
as $\phi(x)$.  
Now, in the B.C.'s
(II) and (III), $\phi$ is left arbitrary 
at $x^-=\pm L$ and so are $\delta\phi(x)$ and $v(x)$,  which implies that 
the B.C.'s
(II) and (III) 
do not generally satisfy the condition (\ref{gc}). 
As to
(I), 
we can show
that the inverse of the Dirac matrix 
$C^{-1}(x, y)=-C^{-1}(y, x)$ such that
$
C^{-1}(\pm L, y)=C^{-1}(x, \pm L)=0
$ 
does not exist
for
$C(x,y)$ such that $C(\pm L,y)=C(x,\pm L)=0$. 
Therefore, besides the periodic B.C.,
the only constraint which may give rise to the
consistent 
theory is 
the anti-periodic B.C..

There is no zero mode in the anti-periodic B.C., since the zero mode is 
constant with respect to $x^-$ and hence cannot satisfy the anti-periodic B.C..
Accordingly, the canonical DLCQ commutator in the case of {\it anti-periodic 
B.C.} takes {\it the same form as
the standard  commutator} (\ref{CCR}) of the continuum theory 
{\it even for the finite} $L$:\cite{TY,Mus}
\begin{eqnarray}
\left[ \phi(x), \phi(y)\right]_{x^+=y^+}= -\frac{i}{4}\epsilon (x^- -y^-)
\delta (x^\bot-y^\bot).
\label{CCR2}
\end{eqnarray}
There is no complication of the zero mode
in this case,
which is a good news for technical reason but a bad news for physics reason.
From the physics point of view, the anti-periodic
B.C.,
though a consistent theory,
 is not interesting, since it forbids zero
mode and hence the vacuum
expectation of the scalar field, 
$\langle 0\vert \phi(x)\vert 0 \rangle =0$,
namely we have no NG phase.

%% file: dpnu973.tex

\section{``Spontaneous Symmetry Breaking'' on the Light Front}
\my
\subsection{Nambu-Goldstone Phase without Nambu-Goldstone Theorem}
Now that the trivial vacuum is established in DLCQ, we are
interested in how it accommodates NG phase which is accounted for by
the complicated vacuum structure in the equal-time quantization.
The trivial vacuum $a(\vec{p},x^+) \vert 0\rangle=0$ ((\ref{trivacDLCQ}))
implies that the vacuum is singlet: 
\begin{eqnarray}
Q \vert 0\rangle=\int d \vec{x} J^+(x) \vert 0\rangle=0,
\label{chargeann}
\end{eqnarray}
where $Q$ is a generic charge which may or may not correspond to
the spontaneously broken symmetry.  
The peculiarity of the LF charge is that {\it even the non-conserved charge
annihilates the vacuum}
(Jers\'ak-Stern theorem)\cite{JS}, which was proved in the absence
of massless particle. It essentially comes from
the $P^+$-conservation due to the integral $\int d x^-$
in (\ref{chargeann})
: The
charge communicates the vacuum ($P^+=0$) to only the state with $P^+=0$,
the zero mode. In DLCQ it is easy to see that
the absence of the zero mode state in the Fock
space of DLCQ implies $\langle A\vert Q\vert 0\rangle=0$ for
any state $\vert A \rangle$ \cite{MY} and hence
$Q\vert 0\rangle =0$. 
This was demonstrated \cite{MY} in the explicit model, the sigma model with
explicit chiral symmetry breaking, which we shall discuss later in detail.
It has further been shown\cite{KTY,TY} that the result remains  valid 
{\it even in the chiral symmetric limit} 
where the massless particle (NG boson) does exist,
which will be explained 
in the next subsections. 

Here we stress essentially different aspect of the realization of the
symmetry of the LF quantization than that of the equal-time quantization. 
The concept of ``spontaneous symmetry breaking'' in the equal-time quantization
is that 
the symmetry of the action
is not realized as it stands in the real world
in such a way that 
the symmetry is manifest at the operator level, i.e., the Hamiltonian
is invariant under the transformation induced by the charge $Q$, 
while it is ``broken'' at the state level, i.e.,
the vacuum is not invariant \cite{BKY}:
\begin{eqnarray}
i\dot{Q} =[Q,H]=0, \qquad  Q\vert 0\rangle\ne 0.
\label{SSB}
\end{eqnarray}
However, this is impossible in the LF quantization, 
because the LF vacuum is trivial
$Q\vert 0\rangle=0$. 
If the Hamiltonian were also invariant, then both the operator and state 
would be symmetric, thereby the real world 
would have to be symmetric, in disagreement with the phenomenon that we 
wish to describe. In order that the real world is not symmetric in
LF quantization, we 
are forced to accept that 
\begin{eqnarray}
i\dot{Q} =[Q,H] \ne 0, \qquad Q\vert 0\rangle=0,  
\label{NGphase}
\end{eqnarray}
 quite opposite to the equal-time case mentioned above.
Importance of the non-conserved LF charge was first stressed by Ida\cite{Ida}. 

One might suspect that in (\ref{NGphase}) there would be no distinction 
between ``spontaneous breaking'' and ``explicit breaking''
\footnote{
~Eq.(\ref{NGphase}) is
essentially different from the explicit symmetry breaking 
in the equal-time quantization; $[Q, H]\ne 0$ {\it and} $Q\vert0\rangle\ne 0$.
Thus (\ref{NGphase}) is really specific to the LF quantization.  
},
since the
symmetry is already broken at operator level.
However, there
actually
exists a prominent feature of the ``spontaneous
symmetry breaking'', namely {\it the existence of massless NG bosons} coupled to
 the ``spontaneously broken'' currents
 , 
 which is 
 in the equal-time quantization
 ensured by Nambu-Goldstone
(NG) theorem with the charge satisfying (\ref{SSB}).
Although we have {\it no NG theorem in the LF quantization}, we do have
an alternative feature, {\it a singular behavior of the zero mode 
of the NG boson field in 
the symmetric limit which 
ensures the existence of the massless NG boson} 
in a quite different manner than
the equal-time quantization \cite{KTY,TY}. 
We shall later demonstrate in DLCQ \cite{KTY,TY}
that the LF charge must in fact satisfy (\ref{NGphase})
instead of (\ref{SSB}) and this violation of the symmetry at operator level
takes place 
only at quantum level (through regularization) like the 
anomaly. 
In this sense the ``spontaneous symmetry breaking'' is
a misnomer in the LF quantization and we may call  it 
``{\it Nambu-Goldstone (NG) phase}'' or ``{\it NG realization}'' of the symmetry,
the symmetry of the action being {\it realized through the existence of 
NG bosons}.

Before discussing the DLCQ argument of the NG phase, let 
us first explain  that
{\it the non-conservation of charge with
the trivial vacuum in LF quantization}, (\ref{NGphase}), 
is not only the logical and academic possibility but also is in fact
{\it realized in the real world} as a physical manifestation, namely the 
celebrated {\it Adler-Weisberger (AW) sum rule}\cite{AW}
whose algebraization is nothing but the chiral $SU(3)_L\times SU(3)_R$ algebra
of LF charge (\ref{chiralalgebra}).
\\

1. Classification Algebra\\
Since the vacuum is
invariant $Q_\alpha\vert 0\rangle=Q_{5\alpha}\vert 0\rangle=0$ and
the field operator has a definite transformation property
$[Q_{(5)\alpha}, a^\dagger_j]\sim (T_\alpha)^{i,j} a^\dagger_j$, 
so does the state $Q_{(5)\alpha} a^\dagger\vert 0\rangle=
[Q_{(5)\alpha}, a^\dagger]\vert 0\rangle \sim T_\alpha
a^\dagger\vert 0\rangle$, 
then we have a definite transformation property of the states in the
physical Fock space (\ref{Fock}).
Then the LF charge algebra may be used to {\it classify} the physical states
in sharp contrast to the equal-time charge $Q_{5\alpha}\vert 0\rangle \ne 0$.
\\

2. Reducible Representation, or Representation Mixing\\
On the other hand, the non-conservation of charge $[Q,H]\ne 0$ is 
actually the {\it necessity} from the physics point of view \cite{Ida,CHKT}.
Using the LSZ reduction formula, the pion-emission-vertex $A(p_A)
\rightarrow  B(p_B) +\pi^\alpha(q)$ may be written as
\begin{eqnarray}
\langle B, \pi^\alpha\vert A\rangle 
&=&i\int d^4 x e^{i qx} \langle B\vert  (\Box +m_\pi^2)\pi^\alpha(x)\vert A\rangle
\nonumber \\
&=&i(2\pi)^4\delta(p_A^- -p_B^- -q^-)\delta(\vec{p}_A -\vec{p}_B-\vec{q}) 
\nonumber \\
& & \cdot \langle B\vert j_{\pi^\alpha}(0)\vert A\rangle,
\label{LSZ}
\end{eqnarray}
where 
$j_{\pi^\alpha} (x)=(\Box +m_\pi^2)\pi^\alpha(x)$ is the source function
of the pion interpolating field $\pi^\alpha(x)$. 
It is customary \cite{Weinberg} to take the collinear-momentum 
frame, $\vec{q}=0$ and $q^-\ne 0$ (not soft momentum). Then the 
pion-emission-vertex at $q^2=0$
reads
\begin{eqnarray}
i(2\pi)^3  \delta(\vec{p}_A -\vec{p}_B) 
\langle B\vert j_\pi^\alpha(0)\vert A\rangle
=
i\int d \vec{x} \langle B\vert  m_\pi^2\pi^\alpha(x)\vert A\rangle
\nonumber \\
=
\frac{1}{f_\pi} i\int d \vec{x} \langle B\vert 
\partial_\mu J_{5\alpha}^\mu
\vert A\rangle
=
\frac{1}{f_\pi}
\langle B\vert [Q_{5\alpha},P^-]\vert A\rangle,
\label{pivertex}
\end{eqnarray}
where we have noted $\int d^4 
x
e^{iq x}\Box \pi^\alpha(x)
\sim q^2 \tilde{ \pi}(q)
=0$ for the massive pion ($m_\pi^2\ne 0$) with $q^2=0$ 
and use has been made of
the PCAC relation:
\begin{eqnarray}
\partial_\mu J_{5\alpha}^\mu=
f_\pi m_\pi^2 \pi^\alpha(x),
\label{PCAC}
\end{eqnarray}
with $f_\pi \simeq 93 {\rm MeV}$ being the pion decay constant.
Thus the non-zero pion-emission vertex is due to non-conservation
 of LF charge:
\begin{eqnarray}
i\langle B\vert j_{\pi^\alpha}(0)\vert A\rangle 
\ne 0 \Longleftrightarrow
 [Q_{5\alpha},P^-]\ne 0,
\end{eqnarray}
which must be the case {\it even in the chiral symmetric limit}
 $m_\pi^2 \rightarrow 0$ (Subtlety of this limit will be fully 
 discussed \cite{KTY,TY} 
 in the
following  subsections). 

Physical relevance of the LF charge $ Q_{5\alpha}$ is that it gives the
current vertex (analogue of the $g_A$ for the nucleon),
\begin{eqnarray}
 \langle B\vert Q_{5\alpha}\vert A\rangle
 =(2\pi)^3 \delta(\vec{q})
 \langle B\vert J_{5\alpha}^+(0) \vert A\rangle
 \equiv (2\pi)^3 
2 p_A^+ \delta(\vec{q})
X_\alpha^{B,A},
\label{picharge}
\end{eqnarray}
where we have defined the reduced matrix element of LF charge 
$X_\alpha^{B,A}$ which coincides with the Weinberg's
$X$ matrix\cite{Weinberg}. 
Note that
\begin{eqnarray}
\langle B\vert [Q_{5\alpha}, P^-]\vert A\rangle=
\frac{m_A^2-m_B^2}{2p_A^+}\langle B\vert Q_{5\alpha}\vert A\rangle,
\label{chargeme}
\end{eqnarray}
where $p_A^- -p_B^-=(m_A^2 -m_B^2)/2p_A^+$.
Then we observe that non-zero current vertex between different-mass states
is also due to the non-conservation of the LF charge:
\begin{eqnarray}
\langle B\vert Q_{5\alpha} \vert A\rangle 
\ne 0 
\quad (m_A^2\ne m_B^2) 
\Longleftrightarrow
 [Q_{5\alpha},P^-]\ne 0.
\label{mixing}
\end{eqnarray}
Actually, the current vertex is related to the pion-emission-vertex 
(analogue of $G_{N N \pi}$ coupling for the nucleon) 
through (\ref{pivertex}) with (\ref{picharge}) and (\ref{chargeme}): 
\begin{eqnarray}
X_{\alpha}^{B,A}= 
\frac{f_\pi}{m_A^2-m_B^2}
i \langle B \vert j_{\pi^\alpha}(0) \vert A \rangle,
\label{X}
\end{eqnarray} 
which is nothing but the generalized Goldberger-Treiman relation.
 
The non-commutativity between the charge and
$P^-$ implies that the physical state
(eigenstate of $P^-$) is not a simultaneous eigenstate of the charge, i.e.,
does {\it not} belong to irreducible representation of the LF charge algebra. 
The physical states are classified into {\it reducible} representations of LF
charge algebra, which is known for long time as 
``representation mixing''\cite{GH,Weinberg,Ida,Melosh,CHKT}
mostly in the infinite-momentum-frame language.
The non-zero current vertex between different mass states 
actually implies that the charge is leaked among the different 
mass eigenstates, another view of the non-conservation of LF charge. 
To summarize 1. and 2. implied by (\ref{NGphase}), 
the LF charge algebra is not regarded as 
a ``symmetry algebra'' but a ``classification
algebra''\cite{Ida}.\\

3. AW sum rule as a physical manifestation of LF chiral
algebra\\
Now, we demonstrate \cite{Ida} 
that the LF chiral charge algebra is nothing but 
the algebraization of the
AW sum rule. To derive the AW sum rule, we usually start with the equal-time 
chiral charge algebra and then take the infinite-momentum frame, which
means that equal-time algebra as it stands is not the algebraization of
the AW sum rule. On the contrary,  we will see that {\it LF charge algebra
by itself is a direct algebraization of the AW sum rule without recourse to
the infinite momentum frame.} Then the success of the AW sum rule implies that
the {\it LF charge algebra is in fact realized in Nature}.

Let us take a part of  the LF chiral charge algebra (\ref{chiralalgebra}), 
$
[Q^5_1+iQ^5_2, Q^5_1-iQ^5_2]=2Q_3=2I_3
$, where $I_3$ is the third component of the isospin,
$Q^5_\alpha$ is the axial-charge with the isospin
indices $\alpha=1,2,3$. 
Sandwiching them by the proton state $\vert p\rangle$ with momentum $p$ and
$I_3=1/2$, we have 
\begin{eqnarray}
\langle p(p)\vert [Q^5_1+iQ^5_2, Q^5_1-iQ^5_2]
\vert p(p)\rangle
&=&\langle p(p)\vert 2I_3\vert p(p)\rangle
=\langle p(p)\vert p(p)\rangle 
\nonumber \\
&=&(2\pi)^3 2p^+ \delta(\vec{0}).
\label{proton}
\end{eqnarray}
Inserting a complete set $\sum_i \vert i\rangle\langle i\vert=1$
inside the commutator, we have 
\begin{eqnarray}
{\rm L.H.S.}&=&
\int d W \sum_i \delta(W-k_i^-)\int\frac{d \vec{k_i}}{(2\pi)^3 2k_i^+}
\nonumber \\
& & 2\left\{
 \langle p(p)\vert (Q^5_1+iQ^5_2)/\sqrt{2}\vert \alpha_i (k_i)\rangle
 \langle \alpha_i (k_i)\vert (Q^5_1-iQ^5_2)/\sqrt{2}\vert p(p) \rangle
\right.
\nonumber \\
 & &
 \left.- \left(Q^5_1+iQ^5_2 \longleftrightarrow 
 Q^5_1-iQ^5_2\right) 
 \right\}
 \nonumber \\
&=&(2\pi)^3 2p^+\delta (\vec{0})
\cdot \left[
g_A^2
\right.
\nonumber \\
&&
\left.
 +
\int d W \sum_{i\ne n} \delta(W-k_i^-)
2 \left(
  \vert X^{\alpha_i,p}_{-}\vert^2
-\vert X^{\alpha_i,p}_{+}\vert^2
\right)
\right],
\label{saturation}
\end{eqnarray}
where we have used (\ref{picharge}) and 
isolated the neutron state $\vert n\rangle$ 
from $\vert \alpha_i\rangle$;
$ X^{n,p}_{-}=g_A/\sqrt{2}$.
Comparing this with (\ref{proton}), we have
the celebrated AW sum rule\cite{AW}:
\begin{eqnarray}
1=g_A^2 +
\frac{f_\pi^2}{\pi}\int d W \frac{4W}{(W^2-
m_N^2)^2}
Im \left(f^{\pi^- p}(W) - f^{\pi^+ p}(W)
\right),
\label{AW}
\end{eqnarray}
with $m_N=m_p=m_n$ being the nucleon mass, 
where 
\begin{eqnarray}
2 Im f^{\pi^{\pm} p}=\sum_i
2\pi\delta(W-k_i^-) \frac{1}{2k_i^+} \vert \langle \alpha_i\vert 
j_{\pi^{\pm}}(0)\vert p\rangle\vert^2,
\end{eqnarray}
with 
$k_i^-=k_i^+=W, \quad k^2=2 W^2$ 
(
$\vert \alpha_i\rangle$
is at rest), 
is the imaginary part of the forward scattering amplitude of
$\pi^\pm p\rightarrow \pi^\pm p$
and we have used (\ref{X}).
Thus we have established that AW sum rule is in fact 
a direct
expression of the LF charge algebra (\ref{chiralalgebra}). 

As seen from the derivation
itself, the integral part 
on the R.H.S. of (\ref{AW}),
$\sim \sum_i (\vert X_{-}^{\alpha_i,p}\vert^2 -
\vert X_{+}^{\alpha_i,p}\vert^2 ) $ 
in (\ref{saturation})
having contributions from $\vert \alpha_i \rangle$
with masses $m_{\alpha_i}\ne 
m_N$, should not be
zero, since the first term $g_A^2\simeq
(1.25)^2$, coming from the neutron which has a degenerate mass with the proton
$m_N$ (as far as only the strong interaction is concerned),
does not match the L.H.S.$=1$. Then the reality of hadron physics again
dictates
leakage of the charge among different mass eigenstates 
(representation mixing)\cite{GH,Weinberg,Ida,Melosh}:
\begin{eqnarray}
\langle B\vert Q_{5\alpha}\vert A\rangle \ne 0,\quad (m_A\ne m_B),
\end{eqnarray} 
which is equivalent to non-conservation of LF charge
 $[Q_{5\alpha}, P^-] \ne 0$, see (\ref{mixing}).

\subsection{Decoupling Nambu-Goldstone Boson in DLCQ ?}
So far we have argued that NG phase in the LF quantization must be
characterized by (\ref{NGphase}) which is quite opposite 
to the equal-time case (\ref{SSB})
but is actually realized in Nature in hadron physics.
Now we discuss how (\ref{NGphase}) is reconciled with the same physics 
as SSB which is due to the complicated vacuum structure in the equal-time 
quantization \cite{KTY,TY}.
Since the trivial vacuum $Q\vert 0\rangle=0$ 
is established in DLCQ \cite{MY}, the same SSB physics can be realized only 
through the operator side and only operator responsible for 
this phenomenon is the zero mode whose dynamics is governed
by the zero-mode constraint (\ref{zmconstraint2}). 
One might then  expect \cite{HKSW,Rob,BPV}
that the non-perturbative vacuum structure in equal-time quantization is 
simply replaced by the solution of the zero-mode constraint.

However, it was found \cite{KTY,TY} in DLCQ that 
direct use of the {\it zero-mode constraint together with the
current conservation} (for which the NG boson should be 
exactly massless) leads to vanishing of the NG boson-emission-vertex and
the corresponding current vertex, namely {\it decoupling of the NG boson},
and hence no NG phase at all (``no-go theorem''\cite{KTY}). 
This implies that {\it solving the zero-mode
constraint does not give rise to the NG phase in the exact symmetric 
case with zero NG boson mass} $m_\pi^2 \equiv 0$, in sharp contrast to the
expectation mentioned above. As it turns out in the next subsection, 
assuming current conservation
(or equivalently conservation of LF charge) in the NG phase actually leads to 
self-inconsistency, namely the above ``no-go theorem'' is false \cite{KTY},
in perfect agreement with our previous statement (\ref{NGphase}); 
$Q\vert 0\rangle=0$ (trivial vacuum)
and $[Q,P^-]= i\int d \vec{x} \partial_\mu J^\mu \ne 0 $ 
({\it non-conservation} 
of LF charge, or {\it non-vanishing} zero mode ($\vec{q}=0$ )
of the {\it current divergence}).

Let us start with the above (false) ``no-go theorem'' which reveals 
the essential nature of the NG phase in the LF quantization.
In order to treat the zero mode
in a well-defined manner, here we use DLCQ explained in Section 2, 
with the continuum
limit $L\rightarrow \infty$ being taken in the end of whole calculations. 
We always understand 
$\int d \vec{x} =\lim_{L\rightarrow \infty} \int_{-L}^{L} d x^- d^{2} x^\perp$.

The NG-boson ($\pi$) emission vertex $A \rightarrow B + \pi$ may be written as 
(\ref{LSZ}), this time with $m^2_\pi\equiv 0$ and hence $j_\pi(x)
=\Box \pi(x)=(2\partial_+\partial_- -\partial_\perp^2)\pi(x)$. 
Then the NG-boson emission vertex {\it vanishes because of the 
 periodic B.C.}, $\pi(x^-=L)=\pi(x^-=-L)$:
\begin{eqnarray}
(2\pi)^3\delta(\vec{p}_A-\vec{p}_B)
\langle B\vert j_{\pi}(0)\vert
A\rangle
=\int d \vec{x} \langle B\vert \Box \pi(x)\vert
A\rangle \nonumber \\
=\int d x^{\bot}\lim_{L\rightarrow\infty}\langle B\vert 
\Bigl(\int^{L}_{-L}dx^{-}2\partial_{-}\partial_{+}\pi\Bigr)\vert A
\rangle=0.
\label{zeropivertex}
\end{eqnarray}
As seen from (\ref{zmconstraint3}),  
the last equality is nothing but a zero-mode 
constraint for the massless field, and hence {\it the zero-mode
constraint itself 
dictates that the NG-boson vertex should vanish}. Thus we have
established that 
the solution of the zero-mode 
constraint, 
whether perturbative or non-perturbative or even exact, does not lead 
to the NG phase at all. 

Another symptom of this disease is the vanishing of the current vertex
(analogue of $g_A$ in the nucleon matrix element).   
When the continuous symmetry is
spontaneously broken, the NG theorem requires that the corresponding 
current $J_{\mu}$ contains  
an interpolating field of the NG boson $\pi(x)$, that is,    
 $J_{\mu}=-f_{\pi}\partial_{\mu}\pi+\widehat
J_{\mu}$, where $f_{\pi}$ is the ``decay constant'' of the NG boson
and $\widehat J_{\mu}$ denotes the non-pole term.
Then,  the {\it current conservation} $0=\partial_{\mu} J^{\mu}
=\partial_{\mu} \widehat J^{\mu}-f_\pi\Box \pi$ leads to
\begin{eqnarray} 
(2\pi)^3 \delta (\vec{q})\, 
\frac{m_{A}^2-m_{B}^2}{2p_A^+} \langle B \vert \widehat J^+(0)
\vert A \rangle 
=
i \langle B \vert 
\int d \vec{x}\, \partial_{\mu} \widehat J^{\mu}(x) \vert A
\rangle 
\nonumber \\
=i f_\pi \langle B \vert 
\int d \vec{x} \Box \pi(x)\vert A
\rangle
=0,
\label{c-vertex} 
\end{eqnarray} 
where 
the integral of 
the {\it NG-boson sector $\Box\pi$ has no contribution on the LF
because of the periodic B.C.} as we mentioned before. 
Thus the current vertex, 
\begin{eqnarray}
 \langle B\vert \widehat Q \vert A\rangle
 =(2\pi)^3 \delta(\vec{q})
 \langle B\vert \widehat J^+(0) \vert A\rangle
 \equiv (2\pi)^3 
2 p_A^+ \delta(\vec{q})
X^{B,A},
\label{currentvert}
\end{eqnarray}
should vanish at $\vec{q}=0\, (q^2=0)$ for $m^2_{A}\ne m^2_{B}$.

This is actually a manifestation of the conservation of the charge
$\widehat Q$
which contains only the non-pole term.
Note that $\widehat Q$ {\it coincides with the full LF charge}
$Q\equiv
\int d \vec{x}\,  J^{+}$, since   
the pole part always drops out of 
$Q$ by the integration on the LF due to the periodic
B.C., $\pi (x^-=L)=\pi(x^-=-L)$, i.e., 
\begin{eqnarray}
Q=\widehat Q - f_\pi \int d \vec{x}\partial_- \pi(x) =\widehat Q.
\label{nopipole}
\end{eqnarray}
Therefore the {\it conservation of} $\widehat{Q} $ {\it inevitably follows
 from the conservation of} $Q$:
\begin{eqnarray}
[\widehat Q, P^-]=[Q, P^-]=0,
\end{eqnarray}
in contradiction to (\ref{NGphase}). This in fact implies vanishing 
current vertex mentioned above.
This is in sharp contrast to the 
charge integrated over usual space $\mbox{\boldmath$x$}=(x^1,x^2,x^3)$ 
in the equal-time quantization 
where the $\pi$ pole term is  {\it not} dropped
by the integration: Namely, 
$
Q
= 
\int d^3\mbox{\boldmath$x$} J^0$ is conserved, 
while  
$\widehat Q
= \int d^3\mbox{\boldmath$x$} \widehat J^0$ is not:
\begin{eqnarray}
Q
\ne \widehat Q;\quad
[Q
,P^0]=0
,\quad 
[\widehat Q
,P^0]\ne 0.
\end{eqnarray}

Since (\ref{nopipole}) implies that $Q$ contains no massless pole,
we have \cite{KTY,TY}
\begin{eqnarray}
Q\vert 0\rangle =\widehat Q \vert 0\rangle =0,
\end{eqnarray}
simply due to the $P^+$ conservation associated with the integration
$\int d \vec{x}$. Thus we have arrived at $Q\vert 0\rangle=0$ but
$[Q,P^-]=0$, in contrast to (\ref{NGphase}) which we wish to realize
in the LF quantization.   

Here we note that the current vertex is to be
defined by the current {\it without
$\pi$ pole} (\ref{currentvert})
but not by (\ref{picharge}) which we used for the discussion of 
AW sum rule.
However, they are trivially the same, (\ref{nopipole}), when  
the NG boson acquires small mass $m_\pi^2\ne 0$ 
due to explicit symmetry breaking, since then the $\pi$-pole term in 
$ J^+$ ($\sim q^+ \tilde \pi (q)$) is automatically 
dropped for the {\it collinear momentum} $\vec{q}=0 \quad (q^2=0)$, 
with $\tilde \pi (q)\sim j_\pi/(q^2- m_\pi^2)$ being not singular at $q^2=0$. 
What we have shown in the above, {\it based on DLCQ} which has no
ambiguity about the zero mode, is that 
(\ref{nopipole}) does {\it hold even in the
exact symmetric case} $m_\pi^2 \equiv 0$, 
which is highly nontrivial,
since usually the massless $\pi$ pole term survives as in the
equal-time charge mentioned above:
$q^2 \cdot j_\pi/q^2 \rightarrow j_\pi \ne 0$ as $q^2 \rightarrow 0$.
Then the whole analysis done for $Q$ in 3.1 should be understood as that  
for $\widehat Q$ including the exact symmetric case.
 Thus the above result of
vanishing $\pi$-emission vertex and
current vertex would invalidate the whole success of AW sum rule and  
 the reality of the hadron physics. 
 
So, what went wrong?  One might use other B.C. than the
periodic one.
In Section 2 we have argued that beside the periodic B.C., only 
the anti-periodic one can be consistent in DLCQ, which however yields no
NG phase because
of obvious absence of the zero mode.  
One might then give up DLCQ and consider the continuum
theory from the onset, in which case, however, we still need to specify
B.C. in order to have a consistent LF theory
\cite{STH} as
 was discussed also in Section 2. The best we can do in the continuum
theory
will be described in Section 5, which,
however, results in another disaster, namely,
the LF
charge does not annihilate the vacuum, thus invalidating the trivial
vacuum as the
greatest advantage of the whole LF
approach. One also might suspect that the finite volume in $x^-$
direction in DLCQ
could be the cause of this NG-boson decoupling, 
since it is well known that NG phase does not occur in the 
finite volume. However, 
we actually take the $L \rightarrow \infty$ limit in the end, and such a
limit 
in fact can realize NG phase as was demonstrated in the equal-time 
quantization in the infinite volume limit of the finite box 
quantization \cite{NJL}. Moreover, 
in the case at hand in four dimensions, 
the transverse directions $x^{\bot}$ 
extend to infinity anyway.
Hence 
this argument is totally irrelevant. 

Therefore the above result is {\it not 
an artifact} of the periodic B.C. and DLCQ but is deeply
connected to the very nature of the LF quantization, namely the zero
mode.

\subsection{Realization of Nambu-Goldstone Phase in DLCQ}
Such a difficulty was in fact overcome \cite{KTY,TY} by regularizing the theory
through introduction of explicit-symmetry-breaking mass
of the NG boson $m_\pi^2$.\footnote{
~The non-conservation of the NG phase charge on the LF
was stressed by Ida\cite{Ida} and Carlitz et al.\cite{CHKT} 
long time ago
in the continuum theory.
Their way to define the LF charge is somewhat similar to that 
given here, namely,
with the explicit mass of NG boson   kept finite in order to
pick up the matrix element of the current 
without NG-boson pole.
However, they discussed it in the continuum theory without consistent
treatment of the B.C. and hence without realizing the zero mode
problem. 
}
From the physics viewpoint, we actually {\it need explicit symmetry breaking} 
anyway in order to single out the true vacuum out of infinitely 
degenerate vacua in SSB phenomenon 
{\it even in the equal-time quantization} where people are accustomed to
discuss it in the exact symmetric case. Therefore this procedure should be 
quite natural from the physics point of view.

The essence of the NG phase with a small
explicit symmetry breaking can well be described by 
the PCAC (\ref{PCAC}).
  From the PCAC relation the current 
divergence of the non-pole term $\widehat J^{\mu}(x)$ reads 
$
\partial_{\mu}\widehat J^{\mu}(x)=f_{\pi}(\Box+m_{\pi}^2)\pi(x)= 
f_{\pi}j_{\pi}(x).
$
Then we obtain 
\begin{equation}
\label{PCACzero}
\langle B \vert \int d\vec{x}\,
\partial_{\mu}\widehat J^{\mu}(x)
\vert A \rangle 
=f_{\pi}m_{\pi}^2
\langle B \vert 
\int d \vec{x}\, \pi(x)\vert A \rangle  \nonumber \\  
=\langle B \vert 
\int d \vec{x}\, f_{\pi} j_{\pi}(x) 
\vert A \rangle, 
\end{equation}
where the integration of the pole term $\Box \pi(x)$ is dropped 
out as before. 
The equality between the first and the third terms
is a generalized Goldberger-Treiman relation, (\ref{X}), with $X^{B,A}/ 2 p_A^+
=\langle B\vert \widehat J^+(0) \vert A\rangle$. 
The second expression of (\ref{PCACzero})
 is nothing but the matrix element of 
the LF integration of the $\pi$ zero mode (with $P^+=0$)
$\omega_{\pi} \equiv \frac{1}{2L}\int_{-L}^{L} dx^- \pi(x)$.
Suppose that $\int d\vec{x}\, \omega_{\pi} (x)
=\int d\vec{x}\, \pi (x)$ is regular when  
$m_{\pi}^2\rightarrow 0$. Then this leads to the ``no-go theorem''
 again. Thus, in order to have non-zero NG-boson emission 
 vertex (R.H.S. of (\ref{PCACzero}))
  as well as non-zero current vertex (L.H.S.) at $q^2=0$, 
 the $\pi$ zero mode
$\omega_{\pi}(x)$ must behave as 
\begin{equation}
\int d \vec{x}\, \omega_{\pi}\sim \frac 1{m_{\pi}^2} 
\quad (m_{\pi}^2 \rightarrow 0).
\label{omega}
\end{equation}

This situation may be clarified when the PCAC relation is written 
in the momentum space:
\begin{equation}
\displaystyle{
\frac{m_{\pi}^2f_{\pi} j_{\pi}(q^2)} {m_{\pi}^2-q^2}=
 \partial^{\mu} J_{\mu}(q) =
\frac{q^2f_{\pi} j_{\pi}(q^2)}{m_{\pi}^2-q^2}
+\partial^{\mu} \widehat J_{\mu}(q). 
}
\label{PCACmom}
\end{equation}
From this we usually obtain
$\partial^{\mu} \widehat J_{\mu}(q) =(m^2_\pi -q^2)f_\pi
j_\pi(q) /
(m_\pi^2 - q^2)=f_\pi  j_\pi(q)
$, {\it irrespectively of the order} of the two limits $q^2 \rightarrow 0$ and
$m_\pi^2 \rightarrow 0$. In contrast, 
what we have done when we reached the ``no-go theorem'' can be
summarized as follows:
We first have set L.H.S of (\ref{PCACmom}) to
zero (or equivalently, assumed implicitly the regular behavior of
$\int d^3\vec{x}\, \omega_{\pi}(x)$) in the symmetric limit, 
in accord with the current conservation $\partial^{\mu} J_{\mu}=0$. Then   
in the LF formalism with $\vec{q}=0 $ $(q^2=0)$, 
the first term (NG-boson pole term) of R.H.S. was also zero
due to the periodic B.C. or the zero-mode constraint.
Thus we arrived at  $\partial^{\mu}\widehat  J_{\mu}(q) =0$.
However, this procedure is
equivalent to playing a 
nonsense game: 
\begin{eqnarray}
1=\lim_{m_{\pi}^2,\,q^{2}\rightarrow 0}
(\frac{m^{2}_{\pi}-q^{2}}{
m^{2}_{\pi}-q^{2}})=0,
\end{eqnarray}
as far as $f_{\pi} j_{\pi} \ne 0$ (NG phase).
Therefore the {\it ``$m_{\pi}^2 =0$ theory'' with vanishing L.H.S. 
is ill-defined on the LF, namely, the ``no-go theorem'' is false}.
The correct procedure should be to 
take the symmetric limit $m_{\pi}^2 \rightarrow 0$
{\it after} the LF restriction $\vec{q}=0$ $ (q^2=0)$
due to the peculiarity of the LF quantization. 
Then  (\ref{omega}) does follow.

This implies that {\it at quantum level} the LF charge $Q=\widehat Q$ 
 is {\it not conserved}, or the {\it current conservation does not hold}
for a particular Fourier component with $\vec{q}=0$ even 
in the symmetric limit: 
\begin{equation}
\dot{Q}=\frac{1}{i}[Q, P^{-}]=\partial^{\mu} J_{\mu}\vert_{\vec{q}=0}=f_{\pi} 
\lim_{m_{\pi}^2\rightarrow 0} m^2_{\pi}\int d\vec{x}\, \omega_{\pi}
\neq 0,
\label{nonconserv}
\end{equation}
in perfect agreement with (\ref{NGphase}). 
Then, {\it in order to solve the zero-mode constraint for NG phase, 
we need to 
include the explicit symmetry breaking} and then take the
symmetric limit $m_\pi^2 \rightarrow 0$ afterward. This will be
done in the next section in a concrete model. 

To summarize, the NG phase in the LF quantization is realized quite 
differently from that in the equal-time quantization.
Although there is
no NG theorem to ensure the existence of massless NG boson coupled
to the charge (\ref{SSB}), $Q\vert 0\rangle \ne 0,\,  [Q,H] =0$,
we do have a 
singular behavior of the zero mode of the NG field 
(\ref{omega}), which ensures existence of massless NG boson coupled to
the charge (\ref{NGphase}), $Q\vert 0\rangle=0, \, [Q,H]\ne 0$.
In this sense the singular behavior of the $\pi$ zero mode may
be considered as a remnant of the symmetry of the action.  

%% file: dpnu974.tex

\section{Sigma Model in DLCQ}
\my
\subsection{The Model}

Let us now demonstrate \cite{KTY,TY} that   
 (\ref{omega}) and (\ref{nonconserv}) indeed take place {\it
 as the solution
 of the constrained zero modes} in the NG phase
 of the $O(2)$ linear sigma model in DLCQ:  
\begin{equation}\label{lag}
{\cal L}=\frac{1}{2}(\partial_{\mu}\sigma)^2+\frac{1}{2}
(\partial_{\mu}\pi)^2-\frac{1}{2}\mu^2 (\sigma^2+\pi^2)-\frac{\lambda}{4}
(\sigma^2+\pi^2)^2 +c\sigma, 
\end{equation}
 where the last term  
 is the {\it explicit breaking} which regularizes the 
NG-boson zero mode.  
In the equal-time quantization the 
NG phase is well described even at the tree-level. 
We should be able 
to reproduce at least the same tree-level result also in the LF quantization, 
by solving the zero-mode constraints. 
   
As in Sect. 2 we can clearly separate
the zero modes (with $P^{+}=0$), $\pi_0
\equiv \frac{1}{2L}\int_{-L}^{L} dx^- \pi(x)$ (similarly for $\sigma_0$),
from other oscillating modes (with $P^{+} \ne 0$),
$\varphi_{\pi}\equiv \pi-\pi_0$ (similarly for $\varphi_{\sigma}$).
The canonical commutation relations for the 
oscillating modes (\ref{varphicomm}) now read  
\begin{equation}
\left[\varphi_i(x),\varphi_j(y)\right]_{x^+=y^+}
=-{i \over 4}\left[\epsilon(x^--y^-)-{x^--y^- 
\over L}\right]\delta (x^{\bot}-y^{\bot}) \delta_{ij}, 
\label{commutator}
\end{equation}
where each index stands for $\pi$ or $\sigma$. 
 
On the other hand, the zero modes 
 are  not independent degrees of freedom but are 
implicitly determined by $\varphi_{\sigma}$ and $\varphi_{\pi}$
through the
zero-mode constraints (\ref{zmconstraint2}):
\begin{eqnarray}
\chi_{\pi}&\equiv&
\frac 1{2L} \int^L_{-L}dx^-
\left[(\mu^2-\partial_{\bot}^2)\pi
+\lambda \pi(\pi^2+\sigma^2)\right]= 0,\nonumber\\
\chi_{\sigma}&\equiv& 
\frac 1{2L} \int^L_{-L}dx^-
\left[(\mu^2-\partial_{\bot}^2)\sigma
+\lambda \sigma (\pi^2+\sigma^2)-c \right]= 0.
\label{pizero}
\end{eqnarray}
Thus the zero modes can be solved away from the physical Fock space which is 
constructed upon the trivial vacuum (\ref{trivacDLCQ}).
Note again that through the equation of motion  
these constraints are equivalent  to 
the characteristic of the DLCQ with periodic B.C., see (\ref{zmconstraint3}): 
$\chi_{\pi}=$\\
$-\frac{1}{2L}\int^L_{-L}dx^-\,2\partial_{+}\partial_{-}\pi
=0,
$ (similarly for $\sigma$)
which we have used to prove the ``no-go theorem'' for the case of
$m_{\pi}^2\equiv 0$. 

Actually, in the NG phase $(\mu^2 < 0)$  the equation of motion 
of $\pi$ reads
$(\Box+m_{\pi}^2)\pi(x)=j_\pi(x)$, where $j_\pi(x)$ is now
explicitly given by
\begin{eqnarray}
 j_{\pi}(x) =
 -\lambda({\pi}^3+\pi
{\sigma}'^2+2v\pi{\sigma}'),
\label{sourse} 
\end{eqnarray}
with ${\sigma}'=
\sigma-v$ and $m_{\pi}^2=\mu^2+\lambda v^2=c/v$, 
and $v\equiv \langle \sigma\rangle$ being the classical 
vacuum solution  determined by  $\mu^2 v
+\lambda v^3 =c$. 
Integrating the above equation of motion over $\vec{x}$, we have
\begin{equation}
\int d \vec{x}\, j_{\pi}(x) -m_{\pi}^2 \int d \vec{x}\, 
\omega_{\pi}(x) 
=\int d \vec{x}\, \Box\pi(x)
=-\int d \vec{x}\, \chi_{\pi}=0,
\label{eqmot}
\end{equation} 
where $\int d \vec{x}\, \omega_{\pi}(x) =
\int d \vec{x}\, \pi(x)$.
Were it not for the singular behavior (\ref{omega}) for the $\pi$ zero mode
$\omega_{\pi}$, we would have concluded 
\begin{eqnarray}
 (2\pi)^3\delta(\vec{0})\,
\langle \pi \vert j_{\pi}(0) \vert \sigma \rangle=
-\langle \pi \vert \int d \vec{x}\, \chi_{\pi}
\vert \sigma \rangle=0
\label{zerospp}
\end{eqnarray}
in the symmetric limit  $m_{\pi}^2 \rightarrow 0$.
Namely, the NG-boson vertex at $q^2=0$ would have vanished, which is
exactly what we called ``no-go theorem''
now related to the zero-mode constraint $\chi_{\pi}$.
On the contrary, direct evaluation of the matrix element of
$j_{\pi}(x)$ in (\ref{sourse})
in the lowest order perturbation yields non-zero result 
even in the symmetric limit $m_{\pi}^2\rightarrow 0$:
\begin{eqnarray}
\langle \pi \vert j_{\pi}(0) \vert \sigma \rangle 
=-2\lambda v \langle \pi \vert \varphi_{\sigma} \varphi_{\pi}\vert 
\sigma\rangle =-2 \lambda v \ne 0 \quad(\vec{q}=0),
\label{nonzerossp}
\end{eqnarray}
which obviously contradicts (\ref{zerospp}).
Thus we have seen that  
naive use of the zero-mode constraints by setting $m_{\pi}^2\equiv 0$ 
leads to the {\it internal inconsistency} in the NG phase.
The ``no-go theorem'' is again false.  

\subsection{Perturbative Solution of the Zero-Mode Constraint}
The same conclusion can be obtained more systematically by {\it solving 
the zero-mode constraints in the perturbation} 
around the classical (tree level) solution to the zero-mode constraints
which is nothing but the minimum of the classical potential:
$v_{\pi}=0$ and $v_{\sigma}= v$,
$c=\mu^2v+\lambda v^3=v m_\pi^2$, 
where we have
divided the zero modes 
$\pi_{0}$ (or $\sigma_{0}$) into classical constant piece $v_{\pi}$
(or $v_{\sigma}$) and operator part $\omega_{\pi}$ (or 
$\omega_{\sigma}$):
$\pi_{0}=v_{\pi}+\omega_{\pi}, 
\sigma_{0}=v_{\sigma}+\omega_{\sigma}.
$
The operator zero modes 
are solved perturbatively \cite{TY} by substituting the expansion 
$\omega_i=\sum_{k=1}\lambda^k \omega_i^{(k)}$ into $\chi_{\pi}$,
$\chi_{\sigma}$  under the Weyl ordering which is a natural
operator ordering \cite{weyl}.
 
The operator part of the zero-mode constraints are 
explicitly written down under the Weyl ordering as follows:
\begin{eqnarray}\label{op-zero2}
(-m_{\pi}^2+\partial_{\bot}^2)\omega_{\pi}
&=&\frac{\lambda}{2L}\int_{-L}^{L}dx^-(\varphi_{\pi}^3
+\varphi_{\pi}\varphi_{\sigma}^2+2v\varphi_{\pi}
\varphi_{\sigma})
+\frac{\lambda}{2L}\int_{-L}^{L}dx^{-}
\nonumber \\
&& \left\{(\omega_{\pi}\varphi_{\pi}^2+\varphi_{\pi}^2\omega_{\pi}
+\varphi_{\pi}\omega_{\pi}\varphi_{\pi})
+\frac{1}{2}(\omega_{\pi}\varphi_{\sigma}^2+\varphi_{\sigma}^2
\omega_{\pi})
\right. \nonumber\\
&+&\left.\frac{1}{2}(\omega_{\sigma}\varphi_{\sigma}\varphi_{\pi}
+\varphi_{\pi}\omega_{\sigma}\varphi_{\sigma}
+\varphi_{\sigma}\omega_{\sigma}\varphi_{\pi}
+\varphi_{\sigma}\varphi_{\pi}\omega_{\sigma})\right\} \nonumber\\ 
&+&\lambda(\omega_{\pi}^3
+\frac{1}{2}\omega_{\pi}\omega_{\sigma}^2
+\frac{1}{2}\omega_{\sigma}^2\omega_{\pi}
+v\omega_{\pi}\omega_{\sigma}
+v\omega_{\sigma}\omega_{\pi}), 
\label{zmop}
\end{eqnarray}
and similarly for $\omega_\sigma$.
Were it not for the explicit breaking $m_\pi^2$, the L.H.S of the
zero-mode constraint (\ref{zmop}) would vanish after integration
$\int d^2 x^\perp$, thus yielding $\lambda$-independent equation
for $\omega_\pi$, which is nonsense. The same trouble happens to
the massless scalar theory in (1+1) dimensions where the L.H.S. 
of (\ref{zmop}) is identically zero, which is
in accordance with the notorious non-existence of the theory due to 
the Coleman theorem\cite{Cm}. Then we again got inconsistency which
corresponds to the previous
(false) ``no-go theorem''.
 
Now, the lowest order solution of the zero-mode constraint $\chi_{\pi}$
 for $\omega_{\pi}$ takes the form:
\begin{equation}
(-m_{\pi}^2+\partial_{\bot}^2)\, \omega_{\pi}^{(1)}
=\frac{\lambda}{2L}\int_{-L}^{L}dx^-(\varphi_{\pi}^3
+\varphi_{\pi}\varphi_{\sigma}^2+2v\varphi_{\pi}\varphi_{\sigma}),
\label{operatorzero}
\end{equation}
(similarly the solution of $\chi_{\sigma}$ for $\omega_\sigma$),
which in fact yields (\ref{omega}) as
\begin{equation}
\lim_{m_{\pi}^2\rightarrow 0} m_{\pi}^2\int d \vec{x}\, \omega_{\pi}^{(1)}
=-\lambda\int d \vec{x}\, (\varphi_{\pi}^3
+\varphi_{\pi}\varphi_{\sigma}^2+2v\varphi_{\pi}\varphi_{\sigma})
\ne 0.
\label{omega2}
\end{equation}
This actually ensures non-zero 
 $\sigma \rightarrow \pi \pi$ vertex through (\ref{eqmot}): 
\begin{eqnarray}
& &(2\pi)^3 \delta(\vec{0}) \langle \pi \vert j_{\pi}(0) \vert \sigma \rangle
=
\lim_{m_{\pi}^2\rightarrow 0} m_{\pi}^2\int d \vec{x}\, 
\langle \pi\vert \omega_{\pi}^{(1)}\vert \sigma\rangle\nonumber \\
& &=-\lambda\int d \vec{x}\, \langle \pi\vert
2v\varphi_{\pi}\varphi_{\sigma}\vert \sigma\rangle
=-2\lambda v (2\pi)^3 \delta(\vec{0}) \ne 0, 
\end{eqnarray}
which agrees with the previous direct evaluation (\ref{nonzerossp})
as it should.
We can easily check (\ref{omega}) {\it to any order of perturbation} \cite{TY}:
$\int d\vec{x} \omega_\pi^{(k)} \sim 1/m_\pi^2$ and hence
\begin{eqnarray}
\int d\vec{x} \omega_\pi 
=\int d\vec{x}\sum_{k=1} \lambda^{k} \omega_\pi^{(k)} \sim 1/m_\pi^2.
\label{singular}
\end{eqnarray}

\subsection{LF Charge in NG Phase}
Let us next discuss the LF charge operator corresponding to the
NG phase current
\begin{eqnarray}
J_{\mu}&=&:\pi \stackrel{\leftrightarrow}{\partial}_\mu \sigma : 
=\widehat J_\mu -v\partial_\mu \pi,\nonumber\\
\widehat J_\mu&=&
: \pi \stackrel{\leftrightarrow}{\partial}_\mu \sigma^\prime:,
\label{acurrent}
\end{eqnarray}
where  $:\quad :$ stands for the normal product as in
(\ref{normalorder}).  
The corresponding LF charge is given by
\begin{eqnarray}
Q=\widehat Q -v \int d \vec{x} \partial_- \pi
=\widehat Q,
\label{acharge}
\end{eqnarray}
where the $\pi$-pole term was
dropped out because of the 
periodic B.C. ($\pi(L)=\pi(-L)$) 
as in (\ref{nopipole}) 
{\it even in the symmetric limit}
where $\pi$ becomes exactly massless, 
so that $Q=\widehat Q$ is well defined even in such a limit. Moreover, 
the zero mode ($x^-$-independent) term in $\widehat Q$
is  killed by the derivative $\partial_-$ and the $P^+$-conservation;
\begin{eqnarray}
\widehat Q=\int d\vec{x}\,
: \partial_{-}\varphi_{\sigma}\varphi_{\pi}-\partial_{-}
\varphi_{\pi}\varphi_{\sigma} :,
\end{eqnarray} 
so that the charge contains at least one annihilation operator 
to the rightmost,
$a_\pi^\dagger a_\sigma\vert 0\rangle=a_\sigma^\dagger  a_\pi
\vert 0\rangle=0$, and hence
annihilates the vacuum:
\begin{equation}
Q \vert  0 \rangle=0.
\label{vac-annih}
\end{equation}
The result was first given\cite{MY} for $m_\pi^2 \ne 0$ in conformity
with the Jers\'ak-Stern theorem\cite{JS} and has recently
been shown\cite{KTY,TY} to be valid 
even in the chiral symmetric limit $m_\pi^2 \rightarrow 0$.
    
This is also consistent with explicit computation of the
commutators\footnote{
~By explicit calculation with a careful treatment of the zero-modes
contribution, we can also show that
$\langle [Q, \sigma]\rangle =
\langle [Q,\pi]\rangle =0$ \cite{TY}.
}
based on (\ref{commutator}): 
\begin{eqnarray}
[Q, \varphi_{\sigma}] =-i \varphi_{\pi} ,
\qquad 
[Q,\varphi_{\pi}] =i
\varphi_{\sigma},
\end{eqnarray}
and hence 
\begin{eqnarray}
\langle [Q, \varphi_{\sigma}]\rangle =-i \langle\varphi_{\pi}\rangle=0 ,
\qquad 
\langle [Q,\varphi_{\pi}]\rangle =i\langle 
\varphi_{\sigma}\rangle=0,
\end{eqnarray}
which are to be compared
 with  those in the usual equal-time case 
 where the SSB charge does not 
annihilate the vacuum, $Q
\vert 0\rangle\ne 0$: 
\begin{eqnarray}
\langle [Q
,\sigma]\rangle
=-i\langle \pi\rangle=0,\qquad
\langle [Q
,\pi]\rangle=i\langle \sigma\rangle\ne 0.
\end{eqnarray}

The PCAC relation is now an operator relation for
the canonical field $\pi(x)$ with $f_{\pi}=v$ in this model;
$
\partial_\mu J^\mu (x) = v m_\pi^2 \pi(x)
$. 
Then, 
(\ref{singular}) 
 ensures 
 \begin{eqnarray}
 [Q,P^-]=[\widehat Q, P^-] \ne 0,
 \end{eqnarray}
 namely  non-zero 
 current vertex, $\langle\pi\vert \widehat J^{+} \vert \sigma\rangle \ne 0$ 
$ (q^2=0)$, in the symmetric limit.
 We thus conclude that  
the regularized zero-mode
constraints (with $m_\pi^2\ne 0$), (\ref{pizero}),
indeed lead to (\ref{singular}) and then 
the non-conservation of the LF charge in the
symmetric limit $m_{\pi}^2\rightarrow 0$:
\begin{equation} 
\dot{Q}=\frac{1}{i}[Q, P^-]= v 
\lim_{m_{\pi}^2\rightarrow 0} m_{\pi}^2
\int d \vec{x}\, \omega_{\pi}\neq 0.
\end{equation}
This can also be confirmed by direct computation of $[Q, P^-]$ through the 
canonical commutator and explicit 
use of the regularized zero-mode constraints \cite{TY}.

Inclusion of the fermion into the sigma model
does not change the above result.\cite{TY}


%% file: dpnu975.tex

\section{Zero-Mode Problem in the Continuum Theory}
\my
In the previous sections we have seen that DLCQ gives a consistent picture
of the NG phase keeping the trivial vacuum and physical Fock space.
Here we compare the DLCQ result with that of 
the conventional continuum theory which
is based on the standard commutator (\ref{CCR}) with the sign 
function defined by (\ref{sign}).  
There are several arguments arriving at (\ref{CCR}). 

First, we assume Wightman
axioms \cite{SW} 
in which case we have a spectral representation
(Umezawa-Kamefuchi-K\"allen-Lehmann representation)
for the commutator function: 
\begin{eqnarray}
\langle 0|[\phi(x), \phi(0)]|0\rangle
&=&i \int_{0}^{\infty} d \mu^2 \rho (\mu^2) \Delta(x;\mu^2),
\nonumber\\
\int_{0}^{\infty} d \mu^2 \rho (\mu^2)&=&1, \quad \rho(\mu^2)\ge 0.
\label{UKKL})
\end{eqnarray}
If one assumes that LF restriction $x^+=0$ of the theory is well-defined
(which turns out to be false in the next section), 
then it follows that
\begin{eqnarray}
\langle 0|[\phi(x), \phi(0)]|0\rangle|_{x^+=0}&=&
i \int_{0}^{\infty} d \mu^2 \rho (\mu^2) \Delta(x;\mu^2)|_{x^+=0}\nonumber \\
&=&-\frac{i}{4}\epsilon(x^-)
\delta (x^{\perp}) ,
\label{commfun}
\end{eqnarray}
since $i\Delta(x;\mu^2)|_{x^+=0}
=-\frac{i}{4}\epsilon(x^-)\delta(x^{\perp})
$ is independent of $\mu^2$.
If one further assume that the commutator
 is c-number, then the standard form (\ref{CCR}) follows.

Another argument is the canonical quantization based on the Dirac's
method in Section 2, this time without box normalization
$-L\leq x^- \leq L$, which would {\it formally} 
yields (\ref{CCR}) \cite{HRT}. However, the
naive canonical quantization \cite{HRT} without specifying 
the B.C. does not literally work, since
as we mentioned before, without specifying 
B.C., the LF theory is inconsistent even in the continuum theory \cite{STH}, 
which is related to
the ambiguity of the inverse matrix $C^{-1}$
of the Dirac bracket (\ref{DM}) due to the zero mode, see Section 2.3. 
Then we must specify  B.C. explicitly.
If one adopts the anti-periodic B.C., $\phi(x^-=-\infty)=-\phi(x^-=\infty)$,
instead of the periodic one, 
then the canonical
commutator actually coincides with (\ref{CCR}) even in
the DLCQ with $\vert x^-\vert \leq L$ \cite{TY}, as we mentioned in 
Section 2, see (\ref{CCR2}). Thus (\ref{CCR}) may be regarded as a smooth
 continuum limit of the DLCQ with the anti-periodic B.C..\footnote{
~As to the periodic B.C., on the other hand, it was stressed in
(\ref{polecr}) that the continuum
limit $L\rightarrow \infty$ of 
the canonical DLCQ commutator (\ref{comm2}) (or (\ref{varphicomm})
for oscillating fields) does not give the same result as 
the continuum one (\ref{CCR}), 
although the commutator as it stands
formally coincides with each other in such a limit.
}

Now, in the continuum theory with (\ref{CCR})
it is rather difficult to remove the zero mode
 in a sensible manner as was pointed out by
Nakanishi and Yamawaki \cite{NY} long time ago: 
The real problem is {\it not a single mode} with 
$p^+\equiv 0$ (which is merely of zero measure and harmless)
but actually the {\it accumulating point} $p^+\rightarrow 0$ as can been
seen from $1/p^+$ singularity  
in the Fourier transform of the 
sign function $\epsilon (x^-)$ (\ref{sign}) in the continuum commutator
 (\ref{CCR}). 
This prevents us 
from constructing even a free theory on the LF 
({\it no-go theorem} \cite{NY}),
which will be discussed in the next section.

In this section
we discuss another zero-mode problem 
of the continuum theory with respect to the NG phase. Namely,
the {\it LF charge in NG phase does not annihilate the vacuum}, 
if we formulate the sigma model
on the LF with careful treatment of the B.C. \cite{TY}.
Even if we pretend to have removed the zero mode in the NG phase, 
{\it as far as the canonical commutator 
takes the form of the sign function} (\ref{sign}),
it inevitably leads to a nontrivial vacuum, namely, 
the LF charge does not annihilate
the vacuum. This in fact corresponds to the difficulty to remove 
the zero mode as the accumulating point mentioned above (in
contradiction 
to a widely spread expectation \cite{Wils}).  

\subsection{Collapse of the Trivial Vacuum}

Wilson et al.\cite{Wils} proposed an approach to construct the
effective LF Hamiltonian in the continuum theory
by ``removing the zero mode'' and then  ``putting the unusual counter
terms'' to compensate the removal of 
the zero mode. They claimed to have demonstrated the validity of this approach
in the sigma model where the NG phase can be treated at tree level.
Here we arrive at impossibility to remove the zero mode \cite{TY}
in contradiction to Wilson et al.\cite{Wils}, if
we formulate the NG phase of the same sigma model {\it by
careful
treatment of the B.C.}.
As we emphasized in 2.3,
the B.C. in the LF quantization contains dynamical
information 
and is crucial to define the theory.
Then we shall demonstrate that it is actually {\it 
impossible to remove the zero mode in the NG phase 
of the continuum theory} in a manner consistent with the trivial vacuum:
The LF charge does not annihilate the vacuum.
The point is that contrary to the periodic case, (\ref{nopipole}),
the {\it NG boson pole in the LF charge $Q$ is not dropped}
out for the case of the anti-periodic B.C. which is in accord
 with the sign function
in the standard commutator (\ref{CCR}). 

Let us 
discuss again the $O(2)$ sigma model (\ref{lag}) 
(this time {\it without} explicit symmetry breaking term, $c\equiv 0$)
 {\it in the continuum theory} 
with the standard commutator
 (\ref{CCR}):
\begin{equation}
[\phi_i(x), \phi_j(y)]_{x^+=y^+}=-\frac{i}{4}\epsilon(x^--y^-)
\delta(x^{\perp}-y^{\perp})\delta_{ij}, 
\label{FSc}
\end{equation}
instead of (\ref{commutator}),
where the sign function  
is defined by the principal value prescription as in (\ref{sign}) 
which has no $p^+\equiv 0$ mode but does have an
accumulating point $p^+ \rightarrow 0$. 

We first look at the transformation property
of the fields $\phi =\pi, \sigma$.
The SSB current is given by (\ref{acurrent})
and the  corresponding LF charge is (\ref{acharge}).
 From the commutation relation (\ref{FSc})
we easily find 
\begin{eqnarray}
&&[Q, \sigma(x)]=-i\pi(x)+\frac{i}{4}[\pi(x^-=\infty)
+\pi(x^-=-\infty)] ,\nonumber\\
&&[Q, \pi(x)]= i\sigma(x)-\frac{i}{4}[ \sigma(x^-=\infty)
+\sigma(x^-=-\infty)] .
\end{eqnarray}

To obtain a
sensible transformation property of the 
fundamental fields, the surface terms must 
vanish as operators:
\begin{equation}
\pi(x^-=\infty)+\pi(x^-=-\infty)
=\sigma(x^-=\infty)+\sigma(x^-=-\infty)=0 .   
\label{FSd}
\end{equation}
However, this condition, {\it anti-periodic B.C.},
means that the zero mode is not allowed to exist 
and hence its classical part, {\it condensate $\langle \sigma \rangle$, 
does not exist} at all. 
Thus we have no NG phase contrary to 
the initial assumption. 

We then seek for a {\it modification of the B.C.}
to save the condensate and vanishing surface term simultaneously.
The lesson from the above argument is that 
we cannot impose the canonical commutation relation  
for the full fields, because then not only the surface term but also 
the zero mode (and hence condensate) are required to vanish 
due to the relation (\ref{FSd}). 
So,  let us 
first separate the constant part or condensate (classical zero mode)
$\langle \sigma \rangle = v$ from $\sigma$,
and then impose the canonical commutation relations for the 
fields without zero modes, $\pi$ and the shifted field $\sigma^\prime
=\sigma-v$,  
which are now consistent with the anti-periodic B.C.,
\begin{equation}
\pi(x^- =\infty)+\pi(x^-
=-\infty)
=\sigma^\prime(x^- =\infty)+\sigma^\prime(x^- =-\infty)=0 , 
\label{FSh}
\end{equation}
instead of (\ref{FSd}). This actually corresponds to the usual
quantization
around the classical NG phase vacuum in the equal-time quantization.  
The constant part $v$ should be understood to be determined 
by the 
minimum of the classical potential
\begin{eqnarray}
V&=&\frac{1}{2}\mu^2(\sigma^2+\pi^2)
        +\frac{\lambda}{4}(\sigma^2+\pi^2)^2 \quad,\nonumber \\
 &=&\frac{1}{2}m_{\sigma^\prime}^2{\sigma^\prime}^2+\lambda v
 \sigma^\prime({\sigma^\prime}^2+\pi^2)
 +\frac{\lambda}{4}({\sigma^\prime}^2+\pi^2)^2,
\label{FSe}
\end{eqnarray}   
where $v=\sqrt{-\mu^2/\lambda}$, $\mu^2<0$ and 
$m_{\sigma^\prime}^2=
2\lambda v^2$. In the renormalization-group approach, the potential  
(\ref{FSe}) appears as an ``effective Hamiltonian" \cite{Wils},
while the same potential can be obtained simply through shifting
$\sigma$ 
to $\sigma^\prime =\sigma-v$.  
The canonical commutation relation for $\sigma$, (\ref{FSc}), 
is now replaced by that for $\sigma^\prime$.

Now that the quantized
fields have been arranged to obey the anti-periodic B.C.,
one might consider that the zero mode has been removed.
It is not true, however, {\it as far as we are using the commutator with
the sign function}
in which the zero mode as an
accumulating point
is persistent to exist.

Let us look at the LF charge $Q=\int d\vec{x} J^0(x)$, with
$J^\mu(x)$ being given by (\ref{acurrent}).
In contrast to (\ref{acharge}), this time 
the $\pi$ pole term is {\it not} dropped out:
\begin{eqnarray}
Q - \widehat Q = -v \int d \vec{x}\, \partial_- \pi(x)
= v\left(\pi(x^-=-\infty)-\pi(x^-=\infty)\right)\ne 0.
\label{poleincharge}
\end{eqnarray}
 The straightforward calculation based on (\ref{FSc}) for
$\pi$ and $\sigma^\prime$ leads to 
\begin{equation}
[Q, \sigma^\prime (x)]=-i\pi(x),\quad
[Q, \pi(x)]=i\sigma^\prime (x)+\frac{i}{2}v,
\label{crpis}
\end{equation}
where the surface terms vanish due to the anti-periodic B.C., (\ref{FSh}),   
for the same reason as before. 
The constant term on the R.H.S. of (\ref{crpis}), coming from the
pole term in (\ref{acurrent}), has its origin in the 
commutation relation (\ref{FSc});
\begin{equation}
\int d \vec{x}\, [\partial_{-}\pi(x), \pi(y)]=\int d \vec{x}\,
\left(-\frac{i}{2}\delta (\vec{x}-\vec{y})\right) =-\frac{i}{2},  
\label{FSccc}
\end{equation} 
which is consistent with (\ref{poleincharge}) and is contrasted with 
(\ref{polecr}) in the DLCQ with periodic B.C..
Now, Eq.(\ref{crpis}) implies 
\begin{equation}
\langle 0 \vert [Q, \pi(x)]\vert 0 \rangle
=i\langle 0 \vert 
\sigma^\prime(x)\vert 0 \rangle+\frac{i}{2}v 
=\frac{i}{2}v\ne 0 . 
\end{equation}
Then we find that the {\it LF charge does not annihilate the vacuum} ,
$Q\vert 0 \rangle \ne 0$,
and we have lost 
the trivial vacuum which is a vital feature of the LF quantization. 
There actually exist infinite number of zero-mode states 
$\vert \alpha \rangle\equiv e^{i\alpha Q}\vert 0\rangle$
such that $P^+\vert \alpha\rangle = e^{i\alpha Q} P^+\vert 0\rangle=0$,
where we have used
$[P^+,Q]=0$ and $\alpha$ is a real number: All these states satisfy the 
``Fock-vacuum condition'' $a(p^+)\vert \alpha\rangle =0$ and 
hence the true unique vacuum cannot be specified by this condition in
contrast to the usual expectation.
This implies that the 
zero mode has not been removed, 
even though the Hamiltonian has been rearranged 
by shifting the field into the one without exact zero mode $p^+\equiv
0$.\footnote{
~Actually, this result also holds in DLCQ
with finite $L$ as far as the {\it anti-periodic B.C.} is imposed for
the fields $\pi(L)=-\pi(-L), \sigma^\prime(L)=-\sigma^\prime(-L)$,
in which case the canonical commutator takes the same form as
(\ref{FSc}), see (\ref{CCR2}). Thus the absence of zero mode in the
shifted  fields via anti-periodic B.C. does not
implies absence of zero mode in the Fock space {\it as far as we
require the NG phase}.
}
  
This is in sharp contrast to DLCQ in Sect.4 
where the surface terms 
do vanish 
altogether thanks to the additional term
$-(x^- - y^-)/L$  (``subtraction of the zero mode") besides the sign
function 
$\epsilon (x^- - y^-)$ in the canonical commutator (\ref{commutator}),
see (\ref{polecr}).

The resulting Hamiltonian via the field shifting 
coincides with the ``effective
Hamiltonian" of Wilson et al.\cite{Wils} which was obtained 
by ``removing the zero mode and
adding unusual counter terms" for it. 
The essential difference of their result\cite{Wils} from ours is that
 they implicitly assumed vanishing surface terms altogether:
\begin{equation}
\phi(x^- =\infty)=\phi(x^- =-\infty)=\pi(x^- =\infty)=\pi(x^-
=-\infty)=0. 
\label{FSi}
\end{equation}
However, it is actually not allowed, 
because it contradicts the commutation 
relation (\ref{FSc}) whose L.H.S. would vanish for $x^-=\infty$ and
$y^- =-\infty$ if we followed (\ref{FSi}), while R.H.S. is obviously 
non-zero. This can also be seen by (\ref{FSccc}) whose L.H.S. would
vanish by (\ref{FSi}), while R.H.S. does not.
Note that the above difficulty of the continuum theory 
is not the artifact of exact symmetric
limit $m_\pi^2\equiv 0$: The situation remains the same even if we introduce the explicit symmetry breaking as we did in DLCQ in Section 4.

To summarize, in the general continuum LF quantization based on the
canonical 
commutation relation {\it with sign function}, the LF charge does {\it
not annihilate}
 the vacuum.
  It corresponds to impossibility to remove the zero
mode as an accumulating point in the continuum theory in a manner
consistent 
with the trivial vacuum. Thus, in the continuum theory 
the greatest advantage of the
 LF quantization, the simplicity of the vacuum, is lost.

\subsection{Recovery of Trivial Vacuum in $\nu$-Theory}
It was then suggested \cite{TY}
that a possible way out of this problem within
the continuum theory would be the ``$\nu$-theory'' proposed by  
Nakanishi and Yamawaki \cite{NY} which essentially removes 
the zero mode contribution
in the continuum theory.
The $\nu$-theory modifies the sign function (\ref{sign}) in the
commutator (\ref{CCR}) into a function which {\it 
vanishes at $x^-=\pm \infty$},
by shaving the vicinity of the zero mode to tame the $1/p^+$ singularity
as $c_\nu |p^+|^{\nu}/p^+ (\nu >0)$, with a constant $c_\nu (>0)$ having  
dimension $m^{-\nu}$ and $c_\nu \rightarrow 1$ as $\nu\rightarrow 0$:
\begin{eqnarray}
\left[ \phi(x), \phi(y)\right]_{x^+=y^+}
= -\frac{i}{4}\epsilon_{(\nu)} (x^- -y^-)
\delta(x^\bot-y^\bot),
\label{modCCR}
\end{eqnarray}
where $\epsilon_{(\nu)} (x^-)$ is a modified sign function defined by
\begin{eqnarray}
\epsilon_{(\nu)} (x^-) &=&\frac{ic_\nu}{\pi}
\int_{-\infty}^{+\infty}\frac{dp^+}{p^+}\vert p^+\vert^\nu
e^{-ip^+x^-}\nonumber \\
&=& \epsilon (x^-) \frac{c_\nu Y_{1-\nu}(\vert x^-\vert)}
{\cos\frac{\pi}{2}\nu},\nonumber\\
Y_\alpha(u)&\equiv& \frac{u^{\alpha-1}\theta(u)}{\Gamma(\alpha)}.
\label{modsign}
\end{eqnarray}
Note that 
$\epsilon_{(\nu)} (x^-)\rightarrow \epsilon(x^-)$ as $\nu \rightarrow 0$
for $\vert x^-\vert <\infty$. 
Since $\epsilon_{(\nu)}(\pm \infty) =0$ in contrast to
$\epsilon (\pm \infty )=\pm 1\ne 0$,
we have 
\begin{eqnarray}
\lim_{x^- -y^-\rightarrow \pm \infty} 
\left[ \phi(x), \phi(y)\right]_{x^+=y^+}
=0,
\label{ssvanish}
\end{eqnarray}
in accordance with 
\begin{equation}
\phi (x^-=\infty)=\phi (x^-=-\infty)=0.
\label{nubc}
\end{equation}

This theory is expected to yield the
NG phase with trivial vacuum in the same way as in DLCQ.
Actually, there 
is no surface term nor
constant term ($\frac{i}{2}v$) in the commutator 
(\ref{crpis}), since the pole term and the surface term
both drop out  because of (\ref{nubc}).
Hence the transformation property of the fields and the trivial
vacuum 
should be both realized. Also, the LF charge conservation is expected to
follow unless we introduce the explicit symmetry
breaking, which is in fact the same situation as in DLCQ. Thus, 
in order to realize the NG phase we could do the same game as DLCQ:
The non-decoupling of NG boson can be realized in a way consistent with
the trivial vacuum $Q\vert 0 \rangle=0$ 
by introducing the
explicit symmetry breaking mass of the NG-boson so as to have
the singular behavior of the zero mode
of the NG boson, (\ref{nonconserv}).


%% file: dpnu976.tex

\section{Zero Mode and Lorentz Invariance}
\my
\subsection{No-Go Theorem - No Lorentz-Invariant LF Theory }

Finally, we should mention that 
there is a
more serious 
zero-mode
problem 
in the 
continuum 
LF theory, namely 
the no-go theorem found by Nakanishi and Yamawaki \cite{NY}.
Indeed the cause of the whole trouble is the sign function 
in the continuum commutator 
(\ref{CCR}) whose Fourier transform yields  (\ref{hocr}) for
its $p^+ >0$ part.
We can compute the two-point Wightman
function on LF, by using the Fourier transform (\ref{FT})
together with (\ref{trivac}) and the commutator (\ref{hocr}): 
\begin{eqnarray}
\left. \langle 0|\phi(x)\phi(0)|0\rangle \right|_{x^+=0}
&=&\int_0^\infty \frac{d p^+ d q^+}{ (2\pi)^6 4p^+ q^+}
\int d p^\perp
d q^\perp e^{-i\vec{p} \vec{x}} 
[ a(\vec{p}), a^\dagger(\vec{q})] \nonumber \\
&=& \frac{1}{2\pi}\int_{0}^{\infty}\frac{dp^+}{2p^+}e^{-ip^+x^-}\cdot 
\delta (x^{\bot}).
\label{WFLF}
\end{eqnarray}
This  is 
logarithmically divergent at $p^+ = 0$ and local
in $x^{\bot}$ and, more importantly, is
{\it independent of the interaction and the mass}.

We can easily check whether or not this result is correct one by
comparing it with the covariant result {\it in the free theory} 
\cite{NYb}
where the theory is 
explicitly solved in all space-time. The two-point Wightman function 
in the free theory is given at any point $x$ by the well-known 
invariant delta function 
$\Delta^{(+)}(x;m^2)$ given in (\ref{deltaplus}) and 
is written as (\ref{WFcov}) in terms of the Hankel
function $K_1$ in the space-like region $x^2 <0$, whose 
LF restriction, 
$x^+=0$, yields (\ref{WFcov2}), 
which is finite (positive definite), nonlocal in $x^{\bot}$ and
{\it dependent on mass}, in obvious contradiction to the above result
(\ref{WFLF}).
Hence, already for the free field the LF quantization fails to
reproduce the Lorentz-invariant theory.
Actually, the latter
Lorentz-invariant result (\ref{WFcov2}) is 
a consequence of the {\it mass-dependent} regularization of $1/p^+$ 
singularity 
at $p^+\rightarrow 0$
by the infinitely oscillating (mass-dependent) 
phase factor $e^{-i(m^2+p_{\bot}^2)/2p^+ \cdot x^+}$
in the integral of (\ref{WFcov}) 
{\it before taking the LF restriction}
 $x^+ = 0$.
The LF quantization, restricting to $x^+=0$ beforehand, in fact
kills such a regularizing factor and leads to a wrong result
(\ref{WFLF}).  
Thus
the LF restriction from the beginning loses all the information of
dynamics 
carried  by the {\it zero mode as the accumulating point}.
This implies that {\it even a free theory does not exist on the LF}
\cite{NY}.\footnote{
~This singularity $\int_0^\infty d p^+/p^+$ 
 reminds us of the Coleman theorem\cite{Cm}
 on the absence of the massless
 scalar boson in 1+1 dimensions.
}

One might suspect that this conclusion could be an artifact of too
formal 
argument and
irrelevant to the actual physics, since one can construct free 
particle states, namely a free Fock space, with the correct spectra,
as far as the momentum space consideration is concerned. 
However, the above result implies that 
{\it 
quantum field on LF is
ill-defined 
as the operator-valued distribution and so is the operator product on
LF.
}
Then it is rather difficult 
to construct a {\it realistic}
LF Hamiltonian (with interaction) in terms of
the products of {\it local fields
 on the same LF 
 }
 in a way consistent with
the Lorentz invariance, which would be a serious problem even for 
practical physicists. 

In fact, the above difficulty also applies to the interacting theory
satisfying 
the Wightman axioms (no-go theorem) \cite{NY},
in which case we have a spectral representation
for the commutator function (\ref{UKKL}), and hence 
(\ref{commfun})  under the assumption that  
LF restriction $x^+=0$ of the theory were well-defined. 
Now, taking
the $p^+>0$ part of the Fourier component of 
 (\ref{commfun}), one would further obtain
exactly the same result as (\ref{WFLF}) for the
two-point Wightman function at $x^+=0$.
On the other hand, the same
Wightman axioms yield the spectral representation also for 
the two-point Wightman function:
\begin{equation}
\langle 0|\phi(x)\phi(0)|0\rangle
=\int_{0}^{\infty} d \mu^2 \rho (\mu^2) \Delta^{(+)}(x;\mu^2), 
\label{sprep}
\end{equation}
whose LF restriction  depends on $\rho(\mu^2)$, since  
$\Delta^{(+)}(x;\mu^2)|_{x^+=0}$ given as (\ref{WFcov2}) does depend
on $\mu^2$. This disagrees with (\ref{WFLF})
derived from the LF commutator
(\ref{commfun}) which was also the consequence of the Wightman axioms.
Thus we have arrived at self-contradiction within the framework of 
Wightman axioms
under the assumption that LF restriction is well-defined.

An immediate way to resolve this trouble would be to define the theory
on the ``near LF'', $x^+\ne 0$, slightly away from the exact LF,
$x^+\equiv
0$, and then take 
the LF limit $x^+ \rightarrow 0$ 
only 
in the end of whole calculation
as in (\ref{WFcov2}). 
In fact such a prescription was first proposed by Nakanishi and Yabuki
\cite{NYb}
in the continuum framework and later by
Prokhvatilov et al. and others \cite{PF} in the
context of DLCQ. 
However, it was noted \cite{NY} 
that 
the price to pay in this approach is {\it non-vanishing vacuum
polarization} 
as in the equal-time quantization and 
hence {\it we must give up the trivial vacuum, or physical Fock space}, 
which is the most important feature of the LF quantization. Then there
is
no advantage of this approach over the equal-time quantization,
concerning the simplicity of the vacuum in non-perturbative studies. 
Indeed, it was
demonstrated more explicitly \cite{PF2} that  the vacuum is 
nontrivial and there 
exists
nontrivial renormalization 
in the LF Hamiltonian in this approach: It is no longer simple 
to solve dynamics compared with the equal-time quantization.

Thus, in spite of its difficulties with the above no-go theorem, we must
take 
the quantization on the exact LF, $x^+\equiv 0$, from the beginning 
in order to keep
the trivial vacuum and physical Fock space.
Actually, the no-go theorem implies that the LF restriction is 
not compatible with the Wightman axioms. 
Therefore, in order to make the  
{\it theory well-defined on the
exact LF,
}
we are forced to 
{\it
give up some of the Wightman axioms, most naturally 
the Lorentz invariance. 
}
Indeed, DLCQ defined on the exact LF is such a theory: The theory itself
explicitly violates the Lorentz invariance for $L<\infty$ 
and {\it never recovers it even in the limit of} 
$L \rightarrow \infty$ \cite{NY}, as we shall see later.
At the sacrifice of the Lorentz invariance, 
the trivial vacuum is in fact realized in DLCQ \cite{MY} as we have seen 
before. The same is true in the $\nu$-theory \cite{NY} 
as 
we discussed in Section 5 and further demonstrate in the following.

\subsection{The $\nu$-Theory}
In the $\nu$-theory
the two-point Wightman function for the free theory is given by
\cite{NY}
\begin{eqnarray}
\Delta_{\nu}^{(+)} (x;m^2)
&=& \frac{c_\nu}{(2\pi)^3}\int_{0}^{\infty}
\frac{dp^+ }{2p^+} |p^+|^\nu \int_{-\infty}^{\infty} dp^{\perp}
e^{-ip^-x^+
-ip^+x^- +ip^{\bot}x^{\bot}}
\nonumber\\
&=&\frac{c_\nu [e^{i\pi/2}(x^+ - i0)]^\nu}{4\pi^2}
(\frac{m}{\rho})^{1+\nu}
K_{1+\nu} (m\rho),
\label{WFnu}
\end{eqnarray}
with
$c_\nu (={\rm const.}) > 0 \quad(c_0=1)$ and
$\rho = [-2(x^+-i0)(x^- -i0)+x_\bot^2]^{\frac{1}{2}}$,
where  the extra factor $c_\nu |p^+|^\nu$ is the regularization of the 
zero-mode singularity $1/p^+$ as was mentioned in Section 5.
The 
previous
non-commutativity 
between the integral of (\ref{WFcov}) and $x^+ \rightarrow 0$ 
is now traded for that between $\nu \rightarrow 0$ and $x^+ \rightarrow
0$.
If we take $\nu \rightarrow 0$ first and then $x^+\rightarrow 0$, 
we can reproduce correct
Lorentz-invariant result (\ref{WFcov2}), which is the same as
the procedure to take the ``near LF'' to the 
LF limit $x^+ \rightarrow 0$ \cite{NYb,PF,PF2}.
If, on the other hand, 
we take $x^+ \rightarrow 0$ first and then $\nu \rightarrow 0$, we arrive at
the non-invariant answer (\ref{WFLF}) again.
Thus the theory itself (operator, Fock space, etc.)
violates Lorentz invariance 
and 
{\it never 
reproduces a 
Lorentz-invariant 
field theory even in the limit} $\nu\rightarrow 0$.
Conversely, the $\nu$-theory is well-defined on the exact LF
at the sacrifice of the Lorentz invariance (a part of Wightman axioms).

\subsection{DLCQ and Poincar\'e Algebra}
We can also expect the same situation as $\nu$-theory
in DLCQ even in the limit $L\rightarrow \infty$:
The theory itself is
not Lorentz-invariant,
since the two-point Wightman function 
in the free theory
 takes the form:
\begin{eqnarray}
\Delta_{\rm DLCQ}^{(+)}(x;m^2)\vert_{x^+=0} &=&\frac{1}{2\pi}
\sum_{n>0} \frac{\pi}{L}\frac{1}{2p_n^+} e^{-ip_n^+ x^-}
\cdot \delta(x^\bot) ,\nonumber\\
p_n^+&=& \frac{n\pi}{L} \quad (n=1, 2, \cdot\cdot\cdot),
\label{WFDLCQ}
\end{eqnarray}
which coincides with (\ref{WFLF}) in the continuum 
limit of $L\rightarrow \infty$ (with 
$p_n^+=n\pi/L={\rm fixed}$),  
again in disagreement with the Lorentz-invariant form (\ref{WFcov2}) 
\cite{NY}.
Note that the {\it sum does not include
the zero mode
} $n=0$, since the zero mode
{\it in the free theory}
vanishes through the zero-mode constraint (\ref{zmconstraint2}).

The situation can also be seen by examining the Poincar\'e algebra
of DLCQ in the continuum limit
$L\rightarrow \infty$ which must be taken after whole calculations.
The Poincar\'e generators are given as usual by
\begin{eqnarray}
P^\mu&=&\int_{-L}^{L} d x^- \int d x^\perp T^{+ \mu},\nonumber\\
M^{\mu\nu}&=&\int_{-L}^{L} d x^- \int d x^\perp\left(x^\mu T^{+ \nu}
-x^\nu T^{+ \mu}\right),\nonumber\\
T^{\mu\nu}&=&
\frac{\partial {\cal L}}{\partial \partial_\mu \phi}\partial^\nu\phi
-g^{\mu\nu} {\cal L},\quad \partial^\mu T_{\mu \nu}=0.
\end{eqnarray}
Using the canonical DLCQ commutator (\ref{comm2}), we can
explicitly compute (up to operator ordering) the commutators \cite{MY}:
\begin{eqnarray}
\left[\phi(x), P^\mu\right]
&=&i \partial^\mu \phi(x),\nonumber\\
\left[\phi(x), M^{\mu\nu}\right]
&=&i {\cal M}^{\mu\nu} \phi(x),\nonumber\\
{\cal M}^{\mu\nu}&=&x^\mu\partial^\nu-x^\nu\partial^\mu,
\end{eqnarray}
for $(\mu,\nu)=(i,j), (i,+)$ ($i,j=1,2$), which correspond to the stability
group of LF. As to $M^{+-}$ and $M^{i-}$, however, we have
\begin{eqnarray}
\left[\phi(x), M^{\mu -}\right]&=&
-i{\cal M}^{\mu -}\phi(x) 
- i \beta\int_{-L}^{L}d x^- \left[
(\mu^2-\partial_\perp^2)\phi+\frac{\partial V}{\partial \phi}\right]
\nonumber\\
&+&\delta^{\mu i}\int_{-L}^{L}d \vec{y}
(1-\partial_- y^-) \partial^i \phi(y)
[\phi(x),\phi(y)],
\end{eqnarray}
where $1-\partial_- y^-=L\delta(y^-=L)+\delta(y^-=-L)\ne 0$
due to the periodic B.C., which implies that 
the extra terms violate the invariance \cite{MY}.
Note that although the boost operator $M^{+-}$ belongs to the
stability group of LF in the continuum theory ($L=\infty$),
the periodic B.C. is changed for {\it finite} $L$
which then persists even in the continuum limit
$L\rightarrow \infty$, and hence the DLCQ with
periodic B.C. is not invariant under the boost.
The situation is somewhat different in the DLCQ with 
{\it anti-periodic} B.C. which was shown \cite{TY} to be
boost-invariant \footnote{
~The DLCQ with {\it anti-periodic B.C.} is thus {\it formally} Lorentz-invariant
{\it in (1+1) dimensions} where no $M^{i-}$ exists, 
in sharp contrast to (3+1) dimensions. However,
in (1+1) dimensions where LF coincides with the light cone, 
even the covariant result corresponding to (\ref{WFcov})
has a light-cone singularity on LF $x^+=0$ (no space-like region
at $x^+=0$) and hence {\it becomes ill-defined}: $\Delta^{(+)}(x;m^2)
=K_0 (m\sqrt{-x^2 +i x^0 \epsilon})/2\pi$ is
logarithmically  divergent in the same as (\ref{WFLF}). Thus 
in (1+1) dimensions, the LF quantization coincides with the covariant
theory {\it at the  sacrifice of the well-definedness of both theories}:
Namely, there exists no
theory not only on the exact LF but also in the limit  
from the near LF \cite{NYb,PF,PF2}, anyway. 
This absence of the 
Wightman function on the light cone is due to 
the same kind of singularity as the 
massless scalar propagator in (1+1) dimensions (Coleman theorem)\cite{Cm}.
}
, although invariance under $M^{i-}$
is not recovered anyway even in the continuum limit 
$L\rightarrow \infty$.\footnote{
~This is compared with the old argument \cite{STY} before advent of
DLCQ which claimed, without consistent 
treatment of B.C., that the continuum LF theory is plagued with 
violation of the Poincar\'e invariance due to the surface
term at $x^-=\pm \infty$.
}

Thus the continuum LF theory is not Poincar\'e-invariant
in perfect agreement with the no-go theorem mentioned above.

\subsection{Recovery of Lorentz Invariance at S Matrix Level}
Now, the real problem is how to {\it recover Lorentz invariance of the
physical
quantity (c-number) like S matrix} which, unlike the Wightman function, 
{\it has no reference to the fixed LF},
{\it even though the theory itself, defined on the fixed exact LF, 
has no Lorentz-invariant limit}.
Indeed,
it was pointed out \cite{NY} that
{\it as far as the perturbation theory is
concerned}, the {\it S matrix coincides} in the limit of $\nu\rightarrow
0$ 
{\it with the 
conventional Feynman rule result} which is Lorentz-invariant, 
{\it with one notable exception, 
namely the 
vanishing
vacuum polarization graph}
due to the modification of the zero-mode contribution.
Note that $\nu\rightarrow 0$ is to be taken {\it after whole calculation},
since the $\nu$-theory is defined on the exact LF only for $\nu>0$ 
(no $\nu=0$ theory exists on the exact LF, as dictated by the no-go theorem). 

In fact, 
the Feynman  propagator of the $\nu$-theory takes the form
\cite{NY}:
\begin{eqnarray}
\Delta_{F,\nu} (x;m^2)
=\frac{ic_\nu }{(2\pi)^4} \int d \vec{p} |p^+|^\nu \int d p^-
\frac{e^{-ip^- x^+ +i\vec{p}\vec{x}} }{2p^-p^+ - p_{\bot}^2 -m^2+i0},  
\end{eqnarray}
which is derived from (\ref{WFnu}).
Then the vacuum polarization graph, 
calculated similarly to (\ref{nzerovp0}) \cite{CM},
does vanish \cite{NY}:
\begin{eqnarray}
\int dp^+ |p^+|^\nu\int d p^- \frac {F(p^+p^-)}{2p^+p^- -m^2 +i 0}
=C \int d p^+ |p^+|^\nu \delta(p^+) = 0 ,
\label{zerovp}
\end{eqnarray}
which is fully consistent with
the previous argument
in Section 5
on the trivial vacuum $Q\vert 0\rangle=0$ in the
$\nu$-theory.
Note that the {\it zero-mode contribution $\delta(p^+)$ in (\ref{nzerovp0})
has been modified by
the extra factor $|p^+|^\nu$ ($\nu>0$) so as to yield zero vacuum 
polarization}.
The vanishing vacuum polarization (\ref{zerovp}) 
is in sharp contrast to the case where we 
take $\nu\rightarrow 0$ beforehand, which is
reduced to (\ref{nzerovp0}) having no $|p^+|^\nu$ factor, 
with its whole contribution coming
from  the zero mode $\int dp^+ \delta(p^+)\ne 0$ \cite{CM}.
This actually corresponds to 
 the  prescription \cite{NYb,PF,PF2} approaching from 
``near LF'' to LF, with $\nu=0$, which yields the 
non-vanishing vacuum polarization  
as in the covariant theory based on
the equal-time quantization. 

On the contrary, {\it all other graphs 
having no $\delta(p^+)$ would be
unaffected by the extra factor $|p^+|^\nu$ and thus 
reproduce the usual Lorentz-invariant 
result in the $\nu\rightarrow 0$ limit}. Thus, as far as the perturbation theory
is concerned, we can reproduce Lorentz-invariant result for the S matrix 
which, differing from the Wightman function,
has no reference to the fixed LF time.

Again we can expect the same situation also in DLCQ.  
Although the {\it theory itself is not Lorentz-invariant}, 
we would  {\it reproduce the Lorentz-invariant result 
for the S matrix
except for the vacuum polarization
} in the continuum limit of $L\rightarrow \infty$,
{\it as far as the perturbation theory is concerned}. 
From (\ref{WFDLCQ}) we obtain the Feynman propagator in DLCQ 
which takes the form \cite{TY}:
\begin{equation}
\Delta_{F,{\rm DLCQ}}(x;m^2)
=i\sum_{n=\pm 1, \pm 2, \cdot\cdot\cdot}\frac{1}{2L} 
\int \frac{d p^\bot d p^-}{(2\pi)^3} 
\frac{e^{-ip^- x^+ -ip_n^+ x^- + i p^\bot x^\bot} }
{2p^-p_n^+ - p_{\bot}^2 -m^2+i0},
\end{equation}
where as in (\ref{WFDLCQ})
{\it the zero mode $n=0$ is not included in the sum}.
When this is used in the 
Feynman rule for the perturbation, 
the {\it absence of 
the zero mode} 
$n=0$ actually
dictates that {\it vacuum polarization graph does vanish} \cite{TY}
similarly to
(\ref{zerovp}):
\begin{equation}
\sum_{n=\pm 1, \pm 2, \cdot\cdot\cdot}\frac{\pi}{L}
\int d p^- \frac {F(p^+p^-)}{2p^+p^- -m^2 +i 0}
=C \sum_{n=\pm 1, \pm 2, \cdot\cdot\cdot} \frac{\pi}{L} \delta (p_n^+)
= 0,
\label{zerovpdlcq}
\end{equation}
which is consistent with the trivial vacuum already established 
\cite{MY} through the zero-mode constraint (see Section 2).  
Note that the continuum limit 
($\L\rightarrow \infty$)
of 
(\ref{zerovpdlcq}) 
obviously
disagree
with
the covariant result (\ref{nzerovp0}) \cite{CM}, i.e., 
$
C\int d p^+ \delta(p^+) \quad (\ne 0). 
$
In contrast, {\it all other graphs 
having no $\delta(p_n^+)$ 
are insensitive to the zero mode 
$n=0$
and hence would 
coincide with 
the covariant result in such a limit}.


%% file: dpnu977.tex

\section{Summary and Discussion}
\my
We have discussed various aspects of the zero-mode problem of
the LF quantization.
The zero mode is the ingredient most crucial to the SSB, or NG phase, 
while it is the most dangerous obstacle to realizing the trivial vacuum and the 
physical Fock space of the LF quantization. Moreover, it is deeply 
connected with the Lorentz invariance of the LF theory.

In order to treat the zero mode in a well-defined manner, 
we adopted DLCQ and 
separated it from all other modes. The canonical formalism of DLCQ 
was fully explained.

We then characterized the peculiar nature of the
LF charge in the context of the SSB, or NG phase, namely
the non-conservation $\dot{Q} \ne 0$ and vacuum annihilation
$Q \vert 0\rangle=0$, which is opposite to the equal-time charge,
$\dot{Q}=0$ and $Q \vert 0\rangle \ne 0$. This peculiarity is actually
realized in Nature as the Adler-Weisberger sum rule.

We further 
have studied how the continuous symmetry breaking in (3+1)-dimensions
is described on the LF within the framework of DLCQ.  We have shown that 
it is necessary to introduce an 
explicit
symmetry-breaking mass 
of the NG boson $m_{\pi}$ in order to realize the NG phase in DLCQ. 
The NG phase is reproduced in the limit of  $m_{\pi}\rightarrow 0$,
where 
the peculiar behavior of the NG-boson zero mode is derived: 
The NG-boson zero mode, when integrated over the LF, must behave as 
$\sim 1/m_{\pi}^2$. This ensures the non-vanishing matrix elements 
associated with the NG boson as an inevitable 
consequence that the LF charge is not conserved or  the current 
conservation breaks down even in the limit of $m_{\pi}\rightarrow
0$, in perfect consistency
 with the non-conservation of LF charge in the general argument. 

Here we emphasize that {\it the NG theorem does not exist on the LF}. 
Instead we found
the singular behavior (\ref{omega}) which in fact {\it 
establishes existence of
the massless NG boson coupled to the current such that
$Q|0\rangle =0$ and $\dot{Q}\ne 0$}, quite analogously to
the NG theorem in the equal-time quantization which proves
existence of the massless NG boson coupled to the current such that
$Q|0\rangle \ne 0$ and $\dot{Q}=0$ (opposite
to the LF case).
Thus the singular behavior of the NG-boson 
zero mode (\ref{omega}) (or (\ref{singular})) may be understood as a 
remnant of the symmetry of the action, an analogue of the NG 
theorem in the equal-time quantization. 

The zero mode problem was also discussed in the continuum theory with
careful
treatment of the B.C.. 
It was demonstrated that as far as the sign function is
used for the commutator, the LF charge
does not annihilate the vacuum in sharp contrast to DLCQ, 
since the zero mode as an accumulating point
cannot be removed by simply dropping the exact zero mode with $p^+\equiv
0$
which is just measure zero. 
We also suggested that the $\nu$-theory
might give a possible way out of this nontrivial vacuum problem 
in the continuum theory and give rise to the same result as that in
DLCQ.  

 As to the Lorentz invariance, we have seen that
the no-go theorem forbids the well-defined LF restriction of the 
Lorentz-invariant field theory 
due to the peculiarity of the zero mode
as an accumulating point in the continuum framework. 
Conversely, the theory  
defined on the exact LF such as DLCQ or $\nu$-theory,
although realizing the trivial vacuum and no vacuum polarization,  
would never recover the Lorentz-invariance
even in the limit of $L\rightarrow \infty$ or $\nu\rightarrow 0$. 
Thus the Lorentz-invariant limit in such a theory 
can only be realized on the c-number 
physical quantity like S matrix which has no
reference to the fixed LF but not on the theory itself (Fock space, operator, 
etc). 

In fact, we have discussed that  
as far as the perturbation theory is concerned, both DLCQ and
$\nu$-theory 
would reproduce the Lorentz-invariant S matrix, while keeping the
vacuum 
polarization absent 
(no zero-mode contribution)
in accordance with the trivial vacuum. 
This was shown {\it through the explicit solution of the perturbative
dynamics} which is based on the interaction picture
with the propagator being given by the free theory
whose solution is completely 
known not only on a fixed LF $x^+=0$ but 
also
on other region $x^+\ne 0$.

However, the real purpose of the LF quantization is 
to solve the dynamics {\it non-perturbatively} in a way much simpler than the
equal-time quantization, based on 
the trivial vacuum and the physical Fock space for the interacting 
Heisenberg field.
Then, in order to reproduce the Lorentz invariance 
without recourse to the perturbation theory,
we actually would {\it need explicit solution of the non-perturbative
dynamics} itself, particularly the zero mode solution.
Thus, recovering the Lorentz invariance is a {\it highly dynamical
issue} in the  LF quantization, the situation being 
somewhat analogous to the lattice gauge
theories.
Then it
remains  
a big challenge for the LF quantization to overcome the no-go theorem in
the 
non-perturbative way. Particularly in DLCQ we 
would 
need to find  the
non-perturbative 
solution to the zero-mode constraint 
which
might play a crucial role
in taking
the continuum limit $L\rightarrow \infty$ 
so
as to recover 
the Lorentz invariance
in the physical quantity (c-number).
Or alternatively, we may consider, for example, 
the explicit non-perturbative solution of DLCQ
(with anti-periodic B.C.) 
in (1+1) dimensions, which is {\it formally} Lorentz-invariant 
(see footnote 13 ) \cite{TY}
and might be useful as the first step 
to demonstrate how the Lorentz invariance
is recovered in such a sense.

Finally, it should be mentioned that remarkably enough,
DLCQ has recently been
applied \cite{Suss} to the M-theory as
Matrix theory\cite{BFSS,Banks}, where the zero-mode
problem has also attracted much attention \cite{DOHP}.

\section{Acknowledgements}
I would like to thank S. Tsujimaru for helpful discussions
on many points given in this lecture and also for collaborations
on Refs.10 and 14 which are the core part of the lecture.
Y. Kim is also acknowledged for collaboration on Ref.14. 
Special thanks are due to M. Taniguchi for useful discussions.  
This work was supported in part by a
Grant-in-Aid for Scientific Research from the
Ministry of Education, Science and Culture (No.08640365).
